\documentclass[aps,prc,twocolumn,superscriptaddress,nofootinbib,floatfix]{revtex4-2}
\usepackage{graphicx,amsmath,amssymb,bm}
\usepackage[colorlinks,linkcolor=blue,citecolor=blue,urlcolor=blue]{hyperref}
\newcommand{\MeV}{\,\mathrm{MeV}}
\begin{document}

\title{Self-consistent orbital-free nuclear density functional theory with a physics-constrained learned nonlocal kinetic energy functional}

\author{Fumihiro Imoto}

\date{July 2026}

\begin{abstract}
Nonlocal kinetic-energy density functionals (KEDFs) can encode nuclear shell
structure in orbital-free density functional theory (OFDFT), but
self-consistency requires accurate functional derivatives and a stable
solution of the Euler equation. We construct a density-dependent-kernel KEDF
whose correction is learned from Kohn--Sham (KS) reference data. Linearity in
the kernel shape yields analytic Euler--Lagrange (EL) responses. The fit uses
exact energy-matching equalities and soft quadratic inequality penalties for
violations of prescribed tail-response and selected-path energy-rise margins,
evaluated by an active-set repeated penalized least-squares iteration. We
formulate the radial EL equation as the rearranged one-orbital eigenproblem. 
In a constant-$k_F$ $^{16}$O benchmark it converges
without density mixing and agrees with imaginary-time evolution (ITE) to
sub-keV energy. Adaptive-step ITE gives final species EL residuals of at most
$0.07\MeV$ in the reported calculations, and rearranged diagonalization
reaches the same stationary densities. For spherical $N=Z$ systems without
spin--orbit or Coulomb terms, nucleus-specific fits for $A=16$--$140$
reproduce shell patterns and radii. A correction trained on three nuclei
transfers the shell pattern and radius, but not the absolute energy, to a
previously unseen $A=140$ system. Preliminary fits without the energy-rise rows collapse to centrally depleted
profiles; including those rows removes the instability at the fit stage,
while a Thomas--Fermi-limit-preserving spatial-gradient term, implemented and
released, neither repairs the preliminary collapse nor is needed by the final
functional. A Gaussian-process
formulation provides finite-grid hyperparameter selection and conditional
parameter-uncertainty bands. These results establish self-consistent nonlocal
nuclear OFDFT in this benchmark and quantify its numerical and transferability
limits.
\end{abstract}

\maketitle

\section{Introduction}
\label{sec:intro}

Nuclear density functional theory in its Kohn--Sham (KS) form
\cite{Hohenberg1964,KohnSham1965,Bender2003} is the standard microscopic
framework for medium-mass and heavy nuclei, but its cost is dominated by the
computation of single-particle orbitals. Orbital-free density functional
theory (OFDFT), which expresses the kinetic energy directly as a functional of
the density, offers in principle a dramatically cheaper alternative and has a
long history in electronic-structure theory
\cite{WangCarter2000,XuReview2024}. In nuclear physics, however,
Thomas--Fermi--type and semiclassical functionals fail to reproduce shell
effects, and this failure was long taken as evidence that orbital-free
approaches are intrinsically blind to shell structure, a view challenged only 
recently by data-driven functionals \cite{WuZhao2021,WuCommPhys2025}

Recently, Wu, Col\`o, Hagino, and Zhao \cite{WuPRL2026} demonstrated that a
\emph{nonlocal} KEDF, built from the Lindhard response of nuclear matter in
close analogy with electronic nonlocal functionals
\cite{WangTeter1992,WGC1999,HuangCarter2010,Constantin2018,MGP2018}, does
encode shell effects: evaluated on KS densities, it reproduces the nucleon
localization function (NLF), an established indicator of shell structure. That
work evaluated the functional on given densities; self-consistent
orbital-free calculations, in which the density itself is obtained by
minimizing the nonlocal functional, were left as an open problem. The
difficulty is generic and well documented in electronic ML-KEDF studies: a
functional accurate in \emph{values} on reference densities may still possess
an inaccurate functional \emph{derivative}, so that its variational minimum
drifts away from the physical density
\cite{Snyder2012,Li2016,Meyer2020,Brockherde2017}. In nuclear physics,
kernel-regression KEDFs trained on density values have been explored
\cite{WuZhao2021}, with self-consistency stabilized by projection techniques,
but without nonlocal shell structure.

In this work we demonstrate that self-consistent orbital-free calculations
with a shell-structured nonlocal KEDF are possible, by combining three
ingredients. (i)~An implementation of a density-dependent-kernel (dd) nonlocal
functional with an analytic derivative of the represented finite-table
nonlocal term, in which the Euler--Lagrange (EL) potential includes all kernel
density dependencies. (ii)~\emph{Functional-derivative-informed fitting}: because the
functional is exactly linear in the dimensionless kernel shape $g$, the EL
potential generated by any kernel-shape basis function can be computed
analytically without finite-difference error; the learned kernel correction
$\delta g$ is then fit simultaneously to NLF values, the analytic per-basis
Euler--Lagrange response of the represented functional, and
total energies. This is the functional-space analogue of Sobolev training
\cite{Meyer2020}. (iii)~\emph{Quadratic inequality penalties} that
discourage behavior invisible to value fits: loss of positivity of the
response ratio in the far tail and energy decrease along prescribed density
changes that had produced central-density or tail artifacts in preliminary
self-consistent calculations.
Here ``functional derivative'' refers to variation with respect to the density;
it is distinct from the spatial density gradient $\nabla\rho_q$ introduced
later in the gradient regularization term.

We work in the benchmark setting of Ref.~\cite{WuPRL2026}: $N=Z$ systems, a
simplified Skyrme (SkP-based \cite{Dobaczewski1984}) interaction without
spin-orbit and Coulomb terms, and spherical symmetry. Without spin-orbit
coupling the magic numbers follow the oscillator sequence
$2,8,20,40,70,\dots$, so that $A=16,40,80,140$ are consecutive doubly-magic
$N=Z$ systems, allowing a clean study from light to heavy nuclei. Within this
setting we show: (a)~per-nucleus learned functionals whose self-consistent
densities closely track the KS density and full NLF shell
structure, with the remaining central-density deviations quantified;
(b)~a \emph{single} learned kernel correction, trained on three nuclei with
$A\le 80$, whose final self-consistent densities remain free of the identified
central-depletion and tail artifacts for all
training nuclei and \emph{predict} the NLF shell structure and radius of the
previously unseen $A=140$ system---a central-depletion instability, exposed by preliminary
fits and removed at the fit stage by the selected-path energy-rise penalty; (c)~a Gaussian-process reformulation providing uncertainty
quantification; and (d)~a rearranged-diagonalization formulation whose
constant-$k_F$ $^{16}$O benchmark reaches the same discrete solution as a
prolonged ITE trajectory. We use the descriptive phrase
``physics-constrained learned functional'' throughout. Mathematically the
scheme is kernel-based supervised regression with analytic functional-derivative
supervision. Each fixed-active-set least-squares subproblem is convex, the
energy-matching equalities are exact, and the two inequality-margin terms are
soft penalties rather than hard constraints.

\section{Framework}
\label{sec:framework}

\subsection{Kohn--Sham reference}

The reference model is the simplified Skyrme energy density functional of
Ref.~\cite{WuPRL2026}, based on the SkP parametrization
\cite{Dobaczewski1984} with spin-orbit and Coulomb terms omitted and $N=Z$.
Spherical KS equations are solved on a radial grid ($n_r=500$; box
$R_{\rm box}=20$~fm and $\ell_{\max}=5$--$6$ for $^{16}$O and $^{40}$Ca,
$R_{\rm box}=22$~fm and $\ell_{\max}=12$ for $A=80$ and $A=140$) to a self-consistency residual
below $10^{-7}\,\mathrm{fm}^{-3/2}$. Specifically, the specieswise density-map
residual is
\begin{equation}
 R_{{\rm KS},q}=\left[\frac{1}{N_q}\int d^3r\,
 (\rho_{{\rm out},q}-\rho_{{\rm in},q})^2\right]^{1/2} .
\label{eq:ksresid}
\end{equation}
The released $n_r=250$ initial densities used by the OFDFT calculations are
the corresponding every-second-point samples of these KS density profiles.
For each nucleus the KS calculation yields the density
$\rho_{{\rm KS},q}(r)$ for species $q$, the corresponding KS kinetic density
$\tau_{{\rm KS},q}(r)$, and the reference energy $E_{\rm KS}$. Every density
in this work is a number density: $\rho_q$ has units fm$^{-3}$ and is
normalized as $4\pi\int_0^{R_{\rm box}}r^2\rho_q(r)\,dr=N_q$. Because all
systems considered here have $N=Z$ and identical neutron and proton profiles,
we often write $\rho_{\rm KS}$ for either species and use
$\rho_{\rm tot}=\rho_n+\rho_p=2\rho_q$ for the total nucleon density.
Here $N_n=N$, $N_p=Z$, and $A=N+Z$.

Shell structure is monitored through the species-resolved nucleon localization
function. Suppressing the species label $q$ only in the following display, we
define the dimensionless response ratio
\begin{equation}
\chi(r) \;=\; \frac{\tau(r)-\tau_{\rm vW}(r)}{\tau_{\rm TF}(r)},
\qquad
{\rm NLF}(r) \;=\; \frac{1}{1+\chi(r)^2},
\label{eq:nlf}
\end{equation}
Here $\tau\equiv\tau_q$ is the kinetic density and has units fm$^{-5}$;
the corresponding physical kinetic-energy density is $C\tau_q$, with
$C=\hbar^2/(2m)$ and $m$ the common nucleon mass used in this $N=Z$
benchmark.
$\tau_{\rm TF}=c_{\rm TF}\rho_q^{5/3}$,
$c_{\rm TF}=\tfrac{3}{5}(3\pi^2)^{2/3}$, and
$\tau_{\rm vW}=|\nabla\sqrt{\rho_q}|^2$ is the von~Weizs\"acker kinetic
density. Thus $\chi$ compares the part of $\tau_q$ beyond the von~Weizs\"acker
term with the local Thomas--Fermi scale, while $\mathrm{NLF}\in[0,1]$ is its
bounded localization diagnostic. The KS
NLF exhibits the characteristic shell oscillations (central dips or peaks and
surface maxima) that the tested local and semilocal approximations do not
reproduce. Two
caveats apply throughout: the NLF is even in $\chi$, so NLF agreement is
equivalent to $\chi$ agreement only where $\chi\ge0$ (our fits therefore
match $\chi$ itself, and small negative-$\chi$ excursions occur only in
the far tail); and since the local kinetic-energy density of a nonlocal
functional is not uniquely defined point by point (it is gauge dependent), the
NLF is used as a diagnostic in the specific convention defined by
Eq.~(\ref{eq:tau}).

\subsection{Nonlocal density-dependent-kernel functional}

For either species $q\in\{n,p\}$, the KEDF kinetic density used to define the
NLF is written as the following functional of that species density:
\begin{align}
\tau_q^{\rm KEDF}[\rho_q](r) \;=\;& \tau_{\rm vW}(r)
+ c_{\rm NL}\,\rho_q^{5/6}(r)\!\int\! d^3r'\,
 t(\bm r,\bm r')\, \nonumber\\
&\times g\big(t^{1/3}|\bm r-\bm r'|\big)
 \rho_q^{5/6}(r') \nonumber\\
&-c_{\rm CORR}\,\rho_q^{5/3}(r),
\label{eq:tau}
\end{align}
The unprimed coordinate $\bm r$ is the observation point and $\bm r'$ is
integrated over all space. The symmetric pair-density variable
$t(\bm r,\bm r') = 3\pi^2\,[\rho_q(r)+\rho_q(r')]/2$ has units fm$^{-3}$ and is the cube of
the symmetrized local Fermi momentum,
$k_F(\bm r,\bm r')=t^{1/3}$ (a heuristic pair-density dependence, not a
re-derivation from the Lindhard response of inhomogeneous matter),
$c_{\rm NL}=\tfrac{96}{125}(3\pi^2)^{2/3}$,
$c_{\rm CORR}=\tfrac{21}{125}(3\pi^2)^{2/3}$ are dimensionless constants, and
$g(y)$ is a dimensionless radial kernel shape normalized as
$\int d^3y\,g(y)=4\pi\int_0^\infty y^2g(y)\,dy=1$ analytically, constructed from the
Lindhard function as in Ref.~\cite{WuPRL2026}. Here $q\in\{n,p\}$ labels the
neutron or proton species, $\rho_q$ is the corresponding number density with
$\int\rho_q\,d^3r=N_q$, and $\tau_q^{\rm KEDF}$ is its KEDF kinetic
density; the dimensionless kernel argument is $y=k_F R$ with
$R=|\bm r-\bm r'|$, and the
subscripts abbreviate the \emph{nonlocal} kernel term (NL) and the local
Thomas--Fermi \emph{correction} (CORR). The latter is a subtractive local term
chosen so that the analytic base functional recovers the uniform
Thomas--Fermi limit. In the numerical
representation the
momentum transform is truncated and the resulting kernel is tabulated on a
finite $y$ interval. This finite transform/table representation has a
cutoff-dependent numerical normalization error and is not renormalized
automatically in the present implementation. The local correction
$c_{\rm CORR}\rho_q^{5/3}$ therefore restores the Thomas--Fermi limit of the
analytic (infinite-table) base functional, whereas a finite table retains a
small cutoff-dependent error. The fitted shift $s$ of this local correction is
fixed by finite-nucleus energy equalities and does not impose an additional
uniform-matter condition. Consequently, the finite-table norm depends on the
momentum cutoff and table extent and is not a universal constant of the
implementation. The kinetic energy is
$T_{\rm KEDF}=C\int d^3r\,\tau_{\rm tot}^{\rm KEDF}$, where
$C\equiv\hbar^2/(2m)=20.73553\,\mathrm{MeV\,fm^2}$ and
$\tau_{\rm tot}^{\rm KEDF}=\tau_n^{\rm KEDF}+\tau_p^{\rm KEDF}$. The
gradient regularization introduced in Eq.~(\ref{eq:gradct}) is a separate
energy contribution, so the full kinetic part is
$T=T_{\rm KEDF}+E_\nabla$ (with $E_\nabla=0$ for the base and per-nucleus
functionals). The total OFDFT energy is
$E=T+E_{\rm int}^{\rm SkP}[\rho_{\rm tot}]$ in the present benchmark, where
$E_{\rm int}^{\rm SkP}$ denotes the interaction part of the simplified SkP
functional rather than the complete KS energy.

For clarity, the KEDF part is species diagonal,
\begin{equation}
 T_{\rm KEDF}[\rho_n,\rho_p]=\sum_{q=n,p}T_q^{\rm KEDF}[\rho_q],\qquad
 \rho_{\rm tot}=\rho_n+\rho_p .
\label{eq:species}
\end{equation}
In the present $N=Z$ setting, $\rho_n=\rho_p=\rho_q=\rho_{\rm tot}/2$ and
$T_{\rm KEDF}=2T_q^{\rm KEDF}$, with
$T_q^{\rm KEDF}=C\int d^3r\,\tau_q^{\rm KEDF}(\bm r)$. The nonlocal contribution
for one species can be written as the ordered double integral
\begin{align}
T_{{\rm NL},q}[\rho_q]
={}& Cc_{\rm NL}\!\int\!d^3r\,d^3r'\,
 \rho_q^a(\bm r)\rho_q^a(\bm r')\nonumber\\[-2pt]
&\times f_1(R;t),
\qquad C\equiv\frac{\hbar^2}{2m},\quad a=\frac56,
\label{eq:tnl}
\end{align}
The exponent $a=5/6$ is the power of each explicit density prefactor,
and $C$ converts the integrated kinetic density to energy. The pair
$(\bm r,\bm r')$ is counted in both orders; there is no additional symmetry
factor because Eq.~(\ref{eq:tnl}) is obtained by directly integrating the
local density in Eq.~(\ref{eq:tau}). The dimensional pair kernel is defined by
\begin{align}
 f_1(R;t)&=t\,g(y),\qquad y=t^{1/3}R=k_FR,\nonumber\\
 t&=k_F^3=\frac{3\pi^2}{2}
 [\rho_q(\bm r)+\rho_q(\bm r')],\qquad
 R=|\bm r-\bm r'| .
\label{eq:f1scale}
\end{align}
Here $f_1$ is the dimensional pair kernel: since $t$ has units fm$^{-3}$ and
$g$ is dimensionless, $f_1$ has units fm$^{-3}$. The exact scaling in
Eq.~(\ref{eq:f1scale}) gives
\begin{equation}
 \frac{\partial f_1(R;t)}{\partial t}
 =g(y)+\frac{y}{3}g'(y),\qquad g'(y)\equiv\frac{dg}{dy}.
\label{eq:f1deriv}
\end{equation}
Here the partial derivative holds the separation $R$ fixed. Introducing the
two density-dependent nonlocal integrals
\begin{align}
 {\cal I}_q(\bm r)&=\int d^3r'\,
 \rho_q^a(\bm r') f_1(R;t),\nonumber\\
 {\cal J}_q(\bm r)&=\int d^3r'\,
 \rho_q^a(\bm r')
 \left[g(y)+\frac{y}{3}g'(y)\right],
\label{eq:ijconv}
\end{align}
separates the variation of the explicit density powers
($\mathcal I_q$) from the variation of the density-dependent pair scale
($\mathcal J_q$). Their units are fm$^{-5/2}$ and fm$^{1/2}$, respectively.
They are not translation-invariant convolutions because $t$ depends on both
density values. The nonlocal Euler--Lagrange (EL) potential is then
\begin{align}
 v_{{\rm NL},q}(\bm r)
 &\equiv\frac{\delta T_{{\rm NL},q}}{\delta\rho_q(\bm r)}\nonumber\\
 &=Cc_{\rm NL}\left[
 \frac53\rho_q^{-1/6}(\bm r){\cal I}_q(\bm r)\right.\nonumber\\[-2pt]
 &\hspace{3.5em}\left.
 +3\pi^2\rho_q^{5/6}(\bm r){\cal J}_q(\bm r)
 \right]\nonumber\\
 &\equiv v_{{\rm dir},q}(\bm r)+v_{{\rm rearr},q}(\bm r).
\label{eq:vnl}
\end{align}
Both $v_{{\rm NL},q}$ and the other $v$'s below have units of energy. The
labels \emph{dir} and \emph{rearr} denote, respectively, the direct variation
of the two explicit density prefactors and the rearrangement term generated by
the density dependence of $t$ and hence of $k_F$. The coefficient
$5/3=2(5/6)$ comes from varying the two explicit density
prefactors in the symmetric integrand. The coefficient
$3\pi^2=2(3\pi^2/2)$ similarly combines the density dependence of $t$ at
the two ends of each pair. Including the von~Weizs\"acker and local
correction contributions, the full kinetic potential is
\begin{align}
 v_{T,q}(\bm r)
 &\equiv\frac{\delta T_q}{\delta\rho_q(\bm r)}\nonumber\\
 &=-C\frac{\nabla^2\sqrt{\rho_q}}{\sqrt{\rho_q}}
 +v_{{\rm NL},q}(\bm r)\nonumber\\
 &\hspace{1.5em}-C\frac53\bar c_{\rm CORR}\rho_q^{2/3}(\bm r)
 +v_{\nabla,q}(\bm r).
\label{eq:vtfull}
\end{align}
Here $\bar c_{\rm CORR}=c_{\rm CORR}+s$, with the dimensionless shift $s=0$
for the base functional. The corresponding local contribution is
$T_{{\rm corr},q}=-C\bar c_{\rm CORR}\int d^3r\,\rho_q^{5/3}$ and
$v_{{\rm corr},q}=-(5C/3)\bar c_{\rm CORR}\rho_q^{2/3}$.
The term $v_{\nabla,q}$ [Eq.~(\ref{eq:gradv})] is the gradient-regularization
potential, which vanishes for $s_2=0$ (the base and
per-nucleus functionals) and is nonzero only for the shared multi-nucleus
functional used in the final transfer test.
Stationarity at fixed neutron and proton particle numbers follows from
$\delta\{E-\sum_q\mu_q\int d^3r\,\rho_q\}/\delta\rho_q=0$ and gives the
specieswise EL equation
\begin{equation}
 v_{T,q}(\bm r)
 +\frac{\delta E_{\rm int}^{\rm SkP}}{\delta\rho_{\rm tot}(\bm r)}=\mu_q .
\label{eq:el_species}
\end{equation}
Here $\mu_q$ is the Lagrange multiplier (chemical potential) for $N_q$.
The Skyrme interaction depends on $\rho_{\rm tot}$, so its species derivative
is written as $\delta E_{\rm int}^{\rm SkP}/\delta\rho_{\rm tot}$. Varying the neutron
and proton densities independently gives Eq.~(\ref{eq:el_species}). If the
$N=Z$ restriction is imposed before the
variation, both the total functional and the particle-number constraint
acquire the same factor of two, which cancels. Equations
(\ref{eq:f1deriv})--(\ref{eq:vnl}) are the exact functional derivative of
the represented finite-table nonlocal term: the implementation evaluates
both $g$ and $g'$ from the same spline. (The von~Weizs\"acker and
Skyrme-gradient terms are discretized with central first differences; the
ITE flow is thus not the exact discrete gradient flow of that
first-difference energy---Appendix~\ref{app:numerics}.) The analytic
derivatives are verified against central finite differences to a relative
accuracy of $\sim10^{-11}$ at optimal step
(Appendix~\ref{app:numerics}). Self-consistent solutions are discussed in
Sec.~\ref{sec:solver}.

\subsection{Self-consistent solution strategy}
\label{sec:solver}

Completing self-consistent calculations with a nonlocal functional of the
form (\ref{eq:tau}) requires a stable realization of its analytic EL
potential. We first formulate the radial EL problem through the kinetic
rearrangement~\cite{Imoto2021}; it gives an
iteration in terms of one auxiliary eigenstate. A density left unchanged by
this update is called a fixed point. Its
quantitative validation below is restricted to the base constant-$k_F$
$^{16}$O functional. The learned density-dependent results use the common
production numerical settings described at the end of this subsection and in
Appendix~\ref{app:solver}.
In the benchmark only, the pair kernel is held independent of the density,
\begin{equation}
 f_1^{(0)}(R)=k_{F,0}^3g(k_{F,0}R),\qquad
 k_{F,0}=0.80\,\mathrm{fm}^{-1}.
\label{eq:constkf}
\end{equation}
Thus the explicit $\rho_q^{5/6}$ factors remain density dependent, but the
pair scale does not: ${\cal J}_q=0$ and the kernel-density rearrangement term
in Eq.~(\ref{eq:vnl}) is absent. The algebraic ``rearrangement'' used by the
eigensolver below is a separate concept. We use $n_\mu$ for the number of
Gauss--Legendre nodes for $\mu_{\rm ang}=\cos\theta$ in the angular part of
the nonlocal integral; it is unrelated to the chemical potential $\mu_q$.

\emph{Rearranged-diagonalization formulation.}
Choose a positive dimensionless rearrangement parameter $\lambda$ and write
the kinetic functional as
$T=\lambda T_{\rm vW}+T_r^{(\lambda)}$, where
$T_r^{(\lambda)}=T-\lambda T_{\rm vW}$ is the remainder treated as a
density-dependent local potential that is held fixed while each auxiliary
eigenproblem is solved. For the present
decomposition,
$T=T_{\rm vW}+T_{\rm NL}+T_{\rm corr}+T_\nabla$, with
$T_\nabla\equiv E_\nabla$, this implies
$\delta T_r^{(\lambda)}/\delta\rho_q
=(1-\lambda)v_{{\rm vW},q}+v_{{\rm NL},q}+v_{{\rm corr},q}+v_{\nabla,q}$
(with $v_{\nabla,q}=0$ for $s_2=0$).
Collecting all terms other than the explicitly propagated
$\lambda T_{\rm vW}$ contribution, define
\begin{equation}
 v_{{\rm rest},q}=v_{{\rm NL},q}+v_{{\rm corr},q}+v_{\nabla,q}
 +\frac{\delta E_{\rm int}^{\rm SkP}}{\delta\rho_{\rm tot}} .
\label{eq:vrest}
\end{equation}
Dividing the rearranged EL equation by $\lambda$ and setting
$u_q(r)=r\sqrt{\rho_q(r)}$, with units fm$^{-1/2}$, removes the first radial derivative from the
spherical Laplacian. With
$v_{{\rm vW},q}=-C\nabla^2\sqrt{\rho_q}/\sqrt{\rho_q}$, the resulting
one-dimensional auxiliary eigenproblem is (all potentials are evaluated at
the current neutron and proton densities)
\begin{multline}
-C\frac{d^2u_q(r)}{dr^2}
 +\frac{1-\lambda}{\lambda}v_{{\rm vW},q}(r)u_q(r)\\[-2pt]
 +\frac{1}{\lambda}v_{{\rm rest},q}(r)u_q(r)
 =\frac{\mu_q}{\lambda}u_q(r),
\label{eq:diagscf}
\end{multline}
with $u_q(0)=u_q(R_{\rm box})=0$ and
$4\pi\int_0^{R_{\rm box}}|u_q(r)|^2\,dr=N_q$. The eigenvalue on the
right-hand side is the scaled chemical potential $\mu_q/\lambda$.
At a fixed point the rearranged terms cancel algebraically, so the continuum
solution is independent of $\lambda$. Numerically, the cancellation holds up
to the stated discretization and tail safeguards. At each density-update iteration $k$ we
diagonalize the tridiagonal frozen radial Hamiltonian, use its normalized
lowest eigenstate to define
$\rho_{0,q}^{(k)}(r)=|u_{0,q}^{(k)}(r)/r|^2$, and optionally mix the input and
output densities. Suppressing $q$ in the iteration formula,
\begin{equation}
 \rho^{(k+1)}=(1-\alpha_{\rm mix})\rho^{(k)}
 +\alpha_{\rm mix}\rho_0^{(k)}.
\label{eq:diagmix}
\end{equation}
Here the subscript $0$ denotes the lowest auxiliary eigenstate and the
superscript $(k)$ denotes the density-update iteration. The regular $r\to0$
limit is used when reconstructing $\rho_{0,q}$. The density-mixing fraction
$\alpha_{\rm mix}\in(0,1]$ gives the unmixed map at $\alpha_{\rm mix}=1$;
smaller values under-relax it.
This is a one-density-orbital EL solver, not a many-orbital KS
diagonalization. Convergence is monitored with the weighted density-map
residual
\begin{equation}
 R_\rho^{(k)}=\left[\int d^3r\,
  \bigl(\rho_0^{(k)}-\rho^{(k)}\bigr)^2\right]^{1/2}.
\label{eq:diagresid}
\end{equation}
for each species (identical in the present $N=Z$ calculations). Because the
measure is $d^3r=4\pi r^2dr$ and the integrand is a squared number density,
$R_\rho$ has units fm$^{-3/2}$.

The implementation shares the convolution, analytic potential, and energy
evaluator with ITE; the comparison therefore tests the nonlinear
density-update algorithm rather than independently reimplementing the
functional.

The frozen auxiliary spectrum is a useful but incomplete convergence
diagnostic. Let $\delta\rho^{(k)}$ denote the deviation of the $k$th input
density from a fixed-point density. Linearizing the density map gives,
schematically,
\begin{align}
 \delta\rho^{(k+1)}={}&
 \left[(1-\alpha_{\rm mix})I+\alpha_{\rm mix}{\cal M}_\lambda\right]
 \delta\rho^{(k)},\nonumber\\
 {\cal M}_\lambda={}&{\cal R}_\lambda{\cal K}_\lambda,
\label{eq:diagjacobian}
\end{align}
where $I$ is the identity map, ${\cal M}_\lambda$ is the Jacobian of the
unmixed density map, and ${\cal R}_\lambda$ is the response of the
normalized lowest-eigenstate density to a perturbation of the frozen
potential. The potential-feedback part is
\begin{equation}
 {\cal K}_\lambda=
 \frac{(1-\lambda){\cal K}_{\rm vW}+{\cal K}_{\rm rest}}{\lambda}
\label{eq:diagkernel}
\end{equation}
where ${\cal K}_{\rm vW}=\delta v_{\rm vW}/\delta\rho$ and
${\cal K}_{\rm rest}=\delta v_{\rm rest}/\delta\rho$ are functional
Hessian kernels evaluated at the fixed point; species indices and radial
integration arguments are suppressed. First-order eigenvector
perturbation theory introduces inverse auxiliary level separations into
${\cal R}_\lambda$, but local convergence is controlled by the spectral
radius (the largest absolute eigenvalue) of the full mixed map in
Eq.~(\ref{eq:diagjacobian}), not by one gap
alone.

Table~\ref{tab:diag} reports the constant-$k_F$ $^{16}$O scan. The naive
$(\lambda,\alpha_{\rm mix})=(1,1)$ map does not converge within 500 iterations. Mixing
at $\lambda=1$, or rearrangement with $\lambda=3$ or 10, restores convergence;
$\lambda=3$ requires the fewest density-update iterations among the tested settings:
41 without mixing. For this run $R_\rho<10^{-8}\,
\mathrm{fm}^{-3/2}$, and the late-iteration geometric mean of
$R_\rho^{(k+1)}/R_\rho^{(k)}$ over iterations 20--40 is approximately
$0.71$. The converged frozen gap is $10.852\MeV$. This is the gap of the
auxiliary one-density-orbital Hamiltonian, not a physical many-orbital
nuclear shell gap. Moreover, the converged $\lambda=1$ mixed case has a
larger auxiliary gap, $11.365\MeV$, so the scan does not support a
gap-only explanation of stability.

\begin{table}[tb]
\caption{Rearranged diagonalization for the base constant-$k_F$ $^{16}$O
benchmark ($n_r=250$, $k_F=0.80\,\mathrm{fm}^{-1}$, $n_\mu=24$). Iteration
counts use $R_\rho<10^{-8}\,\mathrm{fm}^{-3/2}$. The gap is between the two
lowest eigenvalues of the final frozen auxiliary Hamiltonian. The dash
denotes no convergence within 500 iterations.}
\label{tab:diag}
\begin{ruledtabular}
\begin{tabular}{ccccc}
$\lambda$ & $\alpha_{\rm mix}$ & iterations & $E$ [MeV] & $\Delta\epsilon_{\rm aux}$ [MeV] \\
\hline
$1$  & $1.0$ & --  & -- & -- \\
$1$  & $0.3$ & $50$  & $-117.1351$ & $11.365$ \\
$3$  & $1.0$ & $41$  & $-117.1351$ & $10.852$ \\
$10$ & $1.0$ & $136$ & $-117.1351$ & $10.411$ \\
\end{tabular}
\end{ruledtabular}
\end{table}

For $(\lambda,\alpha_{\rm mix})=(3,1)$ the diagonalization energy is
$-117.1351065\MeV$. The energy after the fixed budget of 8000 ITE steps is higher by
$41.616$~keV. Extending the same ITE trajectory to 20000 steps reduces the
energy difference to $0.368$~keV and the relative radial $L^2$ density
difference,
$\|\rho_{\rm ITE}-\rho_{\rm diag}\|_2/\|\rho_{\rm diag}\|_2$ with the
$d^3r$ measure, to $4.59\times10^{-4}$. The independently monitored EL
residual at the converged diagonalization density is $0.048\MeV$; this nonzero
value is consistent with the different radial operators used by the map and
by that diagnostic (Appendix~\ref{app:numerics}). Thus the comparison establishes
agreement of two density-update algorithms for the released discretization, not
an exact zero-residual solution.

\emph{Production ITE calculation.}
All learned density-dependent results below are obtained from the KS
densities with the split-step ITE of Appendix~\ref{app:solver} using an
adaptive step-size schedule: the step begins at $\Delta t=10^{-5}\,\mathrm{MeV}^{-1}$ and
is grown over intervals of decreasing monitored energy and shrunk (with a
return to the last accepted state, hereafter rollback) whenever the monitored
energy would increase, up to the maximum step
$\Delta t_{\max}=5\times10^{-5}\,\mathrm{MeV}^{-1}$. With
rollback the monitored energy is non-increasing at the accepted monitor
points by construction. We assess stationarity with the species amplitude
$\phi_q=\sqrt{\rho_q}$, the frozen EL Hamiltonian, its Rayleigh quotient, and
the residual
\begin{align}
 \hat h_q&=-C\nabla^2+v_{{\rm loc},q},\nonumber\\
 \mu_q&=\frac{\int d^3r\,\phi_q\hat h_q\phi_q}
              {\int d^3r\,\phi_q^2},\nonumber\\
 R_{{\rm EL},q}&=\left[4\pi\int_0^{R_{\rm box}}dr\,r^2
 \left|\hat h_q\phi_q-\mu_q\phi_q\right|^2\right]^{1/2} .
\label{eq:elresid}
\end{align}
Here $v_{{\rm loc},q}=v_{{\rm NL},q}+v_{{\rm corr},q}+v_{\nabla,q}
+\delta E_{\rm int}^{\rm SkP}/\delta\rho_{\rm tot}$ contains every
multiplicative term, while $-C\nabla^2$ is applied as the kinetic operator.
Because $\int\phi_q^2d^3r=N_q$, the denominator in the Rayleigh quotient is
$N_q$. The reported $R_{{\rm EL},q}$ is the unnormalized, specieswise norm; it
has units MeV and retains an approximate $\sqrt{N_q}$ size scaling.
A fixed small step relaxes the light nuclei fully but
leaves the heavy systems short of stationarity within the iteration budget;
the adaptive schedule reaches $R_{{\rm EL},q}=0.02$--$0.05\MeV$ for all systems
(Table~\ref{tab:pernucleus}) at the same budget, and the rearranged
diagonalization independently reaches the same fixed points: for the universal
functional the final ITE and diagonalization densities agree to a relative radial
$L^2$ density difference of $2.2\times10^{-5}$, $7.9\times10^{-5}$,
$1.9\times10^{-4}$ and $2.9\times10^{-4}$ for $^{16}$O, $^{40}$Ca, $A{=}80$ and
$A{=}140$, with identical central densities to four digits. These fixed points
are not artifacts of the KS starting profiles: a diagonalization run started
from a Woods--Saxon density unrelated to the KS solution relaxes to the same
$^{16}$O density (relative $L^2$ difference $1\times10^{-7}$). For $A{=}80$
the Woods--Saxon start likewise gives a centrally filled solution, but not
identically the KS-initialized one ($\rho_c=0.197$ versus $0.183$, relative
$L^2$ difference $1.5\times10^{-2}$): as with the angular quadrature
(Appendix~\ref{app:numerics}), the $A{=}80$ central density carries a
percent-level solver-path sensitivity that the lighter systems do not. The
split-step algorithm, residuals, and this initialization test are documented in
Appendix~\ref{app:solver}.

\section{Learning the kernel correction}
\label{sec:method}

\subsection{Parametrization and analytic functional-derivative data}

The base functional of Eq.~(\ref{eq:tau}) reproduces the KS NLF only
qualitatively. We therefore learn a correction to the kernel shape,
\begin{equation}
g(y)\;\to\; g(y) + \delta g(y),\qquad
\delta g(y) = \sum_b c_b\,\tilde\gamma_b(y),
\label{eq:dg}
\end{equation}
together with a fitted shift of the local correction,
$c_{\rm CORR}\to c_{\rm CORR}+s$. The basis contains $B=14$
displaced Gaussians in $y$. Each is Thomas--Fermi neutralized according to
\begin{equation}
 4\pi\int_0^\infty dy\,y^2\widetilde\gamma_b(y)=0,
\label{eq:basisneutral}
\end{equation}
so it does not change the uniform-density contribution when $s$ is held
fixed. The integer $b=1,\ldots,B$ labels a basis function, the amplitudes
$c_b$ and the shift $s$ are dimensionless fit parameters, and $\delta g$ is
dimensionless like $g$. Because
Eq.~(\ref{eq:tau}) is \emph{exactly linear} in $g$, three consequences
follow. First, the kernel correction inherits the full variational machinery
with no new derivative code: Eqs.~(\ref{eq:f1deriv})--(\ref{eq:vnl}) remain
valid under $g\to g+\delta g$ and $g'\to g'+\delta g'$, while
Eq.~(\ref{eq:vtfull}) uses
$\bar c_{\rm CORR}=c_{\rm CORR}+s$. Second, define the complete EL potential
$v_{{\rm EL},q}=v_{T,q}+\delta E_{\rm int}^{\rm SkP}/\delta\rho_{\rm tot}$.
On a fixed KS density its coefficient responses are
\begin{align}
 v_b(r)&=\frac{\partial v_{{\rm EL},q}(r)}{\partial c_b},\nonumber\\
 v_s(r)&=\frac{\partial v_{{\rm EL},q}(r)}{\partial s}
       =-\frac{5C}{3}\rho_q^{2/3}(r),
\label{eq:vbasis}
\end{align}
and have units MeV. They are obtained from analytic EL evaluations, with no
finite-difference approximation to the functional derivative. Third, all
responses used below are affine in the coefficients, so each fixed-active-set
least-squares subproblem is convex.

\subsection{Components of the fitting objective}

Collect the kernel coefficients and local-correction shift in
$\bm\theta=(c_1,\ldots,c_B,s)^{\rm T}$. They are determined from three
least-squares data groups, one exact energy-matching equality per training
nucleus, and two soft inequality-margin penalties, all evaluated on fixed KS
density profiles. These groups are called ``blocks'' in the numerical
implementation.
\begin{enumerate}
\item \emph{NLF-value data}: the response ratio $\chi$ of Eq.~(\ref{eq:nlf}),
which is linear in $\delta g$, is matched to $\chi_{\rm KS}$ with a weight
containing the absolute NLF sensitivity
$S_{\rm NLF}=|d\,{\rm NLF}/d\chi|
=2|\chi_{\rm KS}|/(1+\chi_{\rm KS}^2)^2$ and an additional radial weight
that emphasizes the nuclear interior. The renamed symbol $S_{\rm NLF}$ avoids
confusion with the nonlocal integral ${\cal J}_q$.
\item \emph{EL-potential data} (functional-derivative or Sobolev term): the
complete model EL potential is matched
to a constant chemical potential on the KS density,
$v_{\rm base}(r) + \sum_b c_b v_b(r) + s\,v_s(r) \approx \mu$,
using the analytic per-basis potentials $v_b$. Here $v_{\rm base}$ contains
the von~Weizs\"acker, uncorrected density-dependent kernel, local correction,
and simplified SkP interaction potentials at $\rho_{\rm KS}$; $\mu$ is a
per-nucleus auxiliary chemical potential in MeV.
\item \emph{Exact energy-matching equality}: let $\tau_{\rm dd}$ be the base
density-dependent-kernel kinetic density on $\rho_{\rm KS}$ and let
$\Delta\tau(\bm\theta)$ be the fitted correction, so that
$\tau_{\bm\theta}=\tau_{\rm dd}+\Delta\tau$. The equality
\begin{equation}
 C\int d^3r\,\Delta\tau(\bm\theta)
 =C\int d^3r\,[\tau_{\rm KS}-\tau_{\rm dd}]
\label{eq:energyequality}
\end{equation}
is imposed once per training nucleus. Thus the corrected kinetic energy on
the fixed KS density equals the KS kinetic energy; because the interaction
functional sees the same $\rho_{\rm KS}$, the corrected total energy there is
$E_{\rm KS}$. The factor $C$ may be cancelled, so the released linear system
stores this equality in integrated kinetic-density units. A bordered
Karush--Kuhn--Tucker (KKT) system imposes it exactly.
\item \emph{Surface--tail data}: $\chi$ is matched in a surface--tail band
($r\in[4.5,8]$~fm for light, $[5.5,9]$~fm for heavy nuclei) with a weight
that limits the influence of the exponentially growing $\chi_{\rm KS}$.
\item \emph{Tail-margin penalty}: on $8\le r<11.5$~fm for light nuclei and
$9\le r<13$~fm for heavy nuclei, a quadratic penalty discourages
$\chi<\chi_{\min}$, with the target margin $\chi_{\min}=5$. It prevents spurious zero
crossings of $\chi$, which would appear as artificial ${\rm NLF}=1$ peaks at
$\rho\lesssim 10^{-7}\,{\rm fm}^{-3}$. The oscillatory long-range response is
inherited from the sharp Fermi surface of the Lindhard kernel; the penalty
moves its sign changes outside the stated diagnostic interval.
\item \emph{Selected-path energy-rise penalty}: trial densities are discrete
points $\rho_{\nu j}=\rho_{{\rm KS},\nu}+\xi_j\delta\rho_{\nu j}$ with
$\xi_j>0$, chosen to remain nonnegative and to conserve the particle number.
They follow density changes that produced central-density artifacts in
preliminary self-consistent calculations. A second quadratic penalty discourages
$\Delta E_{\nu j}<\varepsilon$, where $\varepsilon=2\MeV$ and
$\Delta E_{\nu j}=E_{\rm model}[\rho_{\nu j}]
-E_{\rm model}[\rho_{{\rm KS},\nu}]$. These finite path-energy differences
supplement the first-order EL-potential data; they should not be interpreted
as direct measurements of a local Hessian.
\end{enumerate}

The last two contributions can now be stated precisely. On the fixed input
densities, both quantities are affine in the fit coefficients:
\begin{align}
 \chi_{\nu i}(\bm\theta)&=\chi^{(0)}_{\nu i}
 +\bm x_{\nu i}^{\rm T}\bm\theta,\nonumber\\
 \Delta E_{\nu j}(\bm\theta)&=d_{\nu j}
 +\bm q_{\nu j}^{\rm T}\bm\theta .
\label{eq:affinepenalty}
\end{align}
Here $i$ labels a radial point in the tail set ${\cal H}_\nu$, $j$ labels a
trial density, $\nu$ labels a training nucleus, and $d_{\nu j}$ is the base
model's path-energy change. With $[x]_+\equiv\max(x,0)$, the conceptual
one-sided forms are
\begin{align}
 {\cal Q}_{T}(\bm\theta)&=\sum_\nu a_{T,\nu}
 \sum_{i\in{\cal H}_\nu}
 [\chi_{\min}-\chi_{\nu i}(\bm\theta)]_+^2,\nonumber\\
 {\cal Q}_{S}(\bm\theta)&=\sum_{\nu,j}a_{S,\nu j}
 [\varepsilon-\Delta E_{\nu j}(\bm\theta)]_+^2 .
\label{eq:ineqpen}
\end{align}
The positive tail weight $a_{T,\nu}$ is dimensionless and
$a_{S,\nu j}$ has units MeV$^{-2}$. Each term is zero when its target margin
is met and rises quadratically with the amount of violation. These are soft
\emph{quadratic inequality penalties}: they discourage but do not prohibit a
violation. In contrast, Eq.~(\ref{eq:energyequality}) is a hard linear
equality. There are no class labels and no classification loss.

The implementation evaluates the soft penalties by an \emph{active-set
repeated least-squares iteration}. From the coefficients at iteration $k$, it forms
the sets of violated inequalities
\begin{align}
 {\cal A}_{T,\nu}^{(k)}
 &=\{i\in{\cal H}_\nu:\chi_{\nu i}(\bm\theta^{(k)})<\chi_{\min}\},
 \nonumber\\
 {\cal A}_{S}^{(k)}
 &=\{(\nu,j):\Delta E_{\nu j}(\bm\theta^{(k)})<\varepsilon\}.
\label{eq:activesets}
\end{align}
Only rows in these sets contribute quadratic residuals to the next solve.
With the selected rows and their normalizations fixed, the resulting convex
quadratic problem is solved together with the exact energy equalities by the
bordered KKT system; the active sets are then recomputed. We therefore refer
to the procedure more fully as an active-set repeated penalized least-squares
iteration with exact linear equality constraints. Because the tail
normalization may change with the active set, the released algorithm is a
sequence of convex equality-constrained quadratic subproblems, not the
minimization of one fixed global quadratic objective. The released GP code
performs 16 updates; its active sets and coefficients cease to change at the
14th update for the released data. The finite-Gaussian-basis production fits
use the same block structure and active-set machinery with 60 damped updates
(damping $1/2$) and mild coefficient/shift ridge regularizers; they are
released, with all settings fixed in the source, as \texttt{kernel\_fit.py},
which regenerates every per-nucleus and shared kernel from the released
reference data. The exact GP normalizations and weights are
given in Appendix~\ref{app:gp}.

The Gaussian data terms admit an exact Gaussian-process interpretation: since
those observations are linear functionals of
$\delta g$, replacing the finite basis by a reduced-rank Gaussian-process
prior gives GP regression in which the observations may be integrals or
derivatives of the kernel rather than only its point values (operator
observations). The soft inequality penalties are added at the
maximum-a-posteriori (MAP) stage. Hyperparameters are selected by a finite-grid
marginal-likelihood comparison, and fixed-active-set inverse-Hessian bands are
obtained for $\delta g$ and fixed-density predictions
(Sec.~\ref{sec:discussion} and Appendix~\ref{app:gp}).

\section{Results I: per-nucleus functional fitting}
\label{sec:results1}

\begin{figure}[tb]
\includegraphics[width=\linewidth]{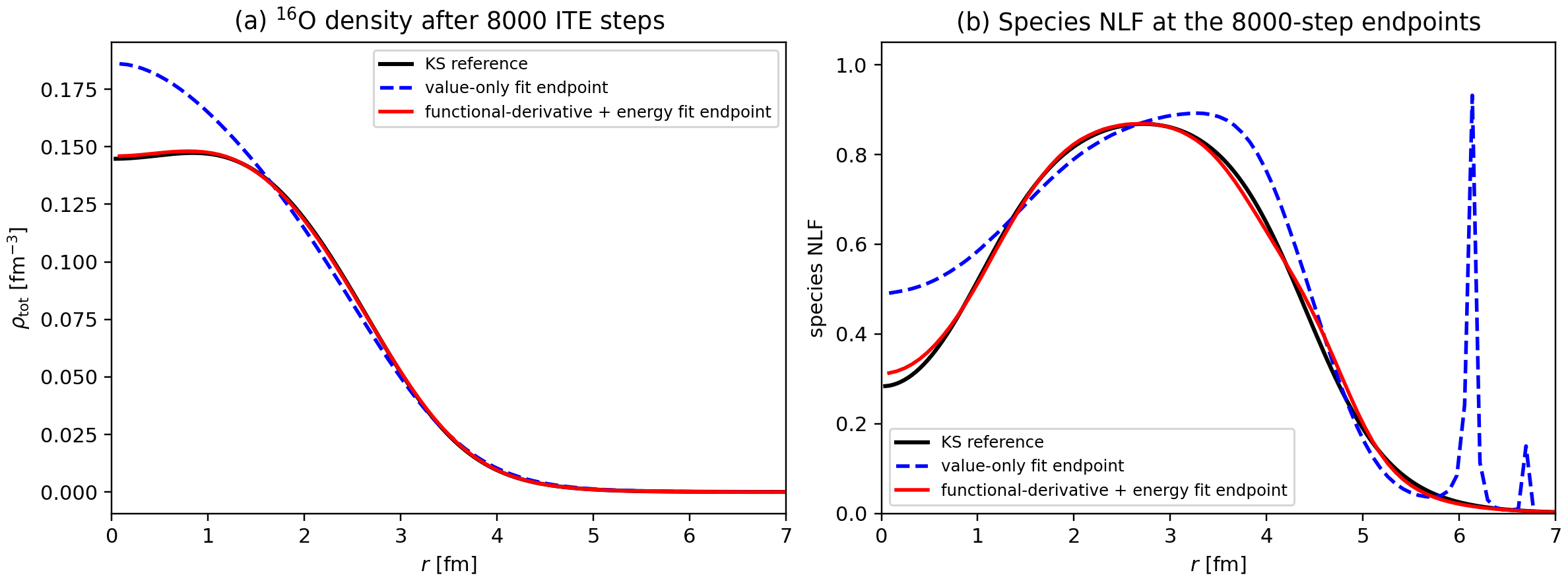}
\caption{Value-only versus functional-derivative-informed learning for
$^{16}$O. (a)~Total nucleon density $\rho_{\rm tot}$: with a value-only fit
(blue dashed) the ITE endpoint moves away from the KS density (black), whereas
adding analytic per-basis EL-potential data and the exact energy equality
(red) keeps it close to $\rho_{\rm KS}$. (b)~The species-resolved NLF at the
corresponding endpoints; neutron and proton NLFs are identical here.}
\label{fig:contrast}
\end{figure}

For the observables reported below, we use the total-density central value
$\rho_c\equiv\rho_{\rm tot}(0)$ and the total-nucleon root-mean-square radius
\begin{equation}
 r_{\rm rms}\equiv
 \left[\frac{1}{A}\int d^3r\,r^2\rho_{\rm tot}(\bm r)\right]^{1/2},
 \qquad A=N+Z .
\label{eq:observables}
\end{equation}
Thus $\rho_c$ has units fm$^{-3}$ and $r_{\rm rms}$ has units fm. In the
present $N=Z$ calculations they can equivalently be evaluated from either
species density after dividing numerator and normalization by two. The NLF
distance $d$ is dimensionless and uses either species, whose neutron and
proton profiles are identical here. We define it explicitly as
\begin{align}
 \Delta_{\rm NLF}(r)&={\rm NLF}_{\rm OF}(r)-{\rm NLF}_{\rm KS}(r),
 \nonumber\\
 d^2&=\frac{\displaystyle\int_0^{R_{\rm NLF}}dr\,
 w_\rho(r)\Delta_{\rm NLF}^2(r)}
 {\displaystyle\int_0^{R_{\rm NLF}}dr\,w_\rho(r)},\nonumber\\
 w_\rho(r)&=r^2\rho_{{\rm KS},q}(r),
\label{eq:nlfdistance}
\end{align}
with $R_{\rm NLF}=7$~fm for $^{16}$O and $^{40}$Ca and 9~fm for the two
heavier systems. The common angular factor $4\pi$ cancels between numerator
and denominator.

Figure~\ref{fig:contrast} isolates the central methodological point. A
$\delta g$ fitted to NLF \emph{values} only reproduces the KS NLF essentially
exactly on $\rho_{\rm KS}$ ($d\simeq10^{-3}$), yet its
self-consistent relaxation drifts by $-2.2\MeV$ ($^{16}$O) and
$-16.4\MeV$ ($^{40}$Ca), the central density overshoots by up to $30\%$, and
the NLF structure at the final density degrades ($d=0.05$--$0.2$). This is the
nuclear manifestation of the value/derivative dichotomy known from electronic
ML-KEDFs \cite{Snyder2012,Li2016,Meyer2020}. The weighted root-mean-square
EL-fit mismatch is
$R_{{\rm fit},\nu}=[\sum_iw_{{\rm EL},\nu i}
(v_{{\rm EL},\nu i}-\mu_\nu)^2/\sum_iw_{{\rm EL},\nu i}]^{1/2}$; it is a
training-density diagnostic and is distinct from the self-consistent residual
$R_{{\rm EL},q}$ in Eq.~(\ref{eq:elresid}). Adding functional-derivative
matching and the exact energy equality reduces $R_{{\rm fit},\nu}$ on
$\rho_{\rm KS}$ from $5.7$ to $0.4\MeV$ for $^{16}$O and keeps the relaxation
near $\rho_{\rm KS}$:
over the converged adaptive-step trajectory the energy moves by only
$0.018\MeV$, and the final density reproduces $\rho_c$ and $r_{\rm rms}$ of the KS
solution to three digits.

\begin{figure}[tb]
\includegraphics[width=\linewidth]{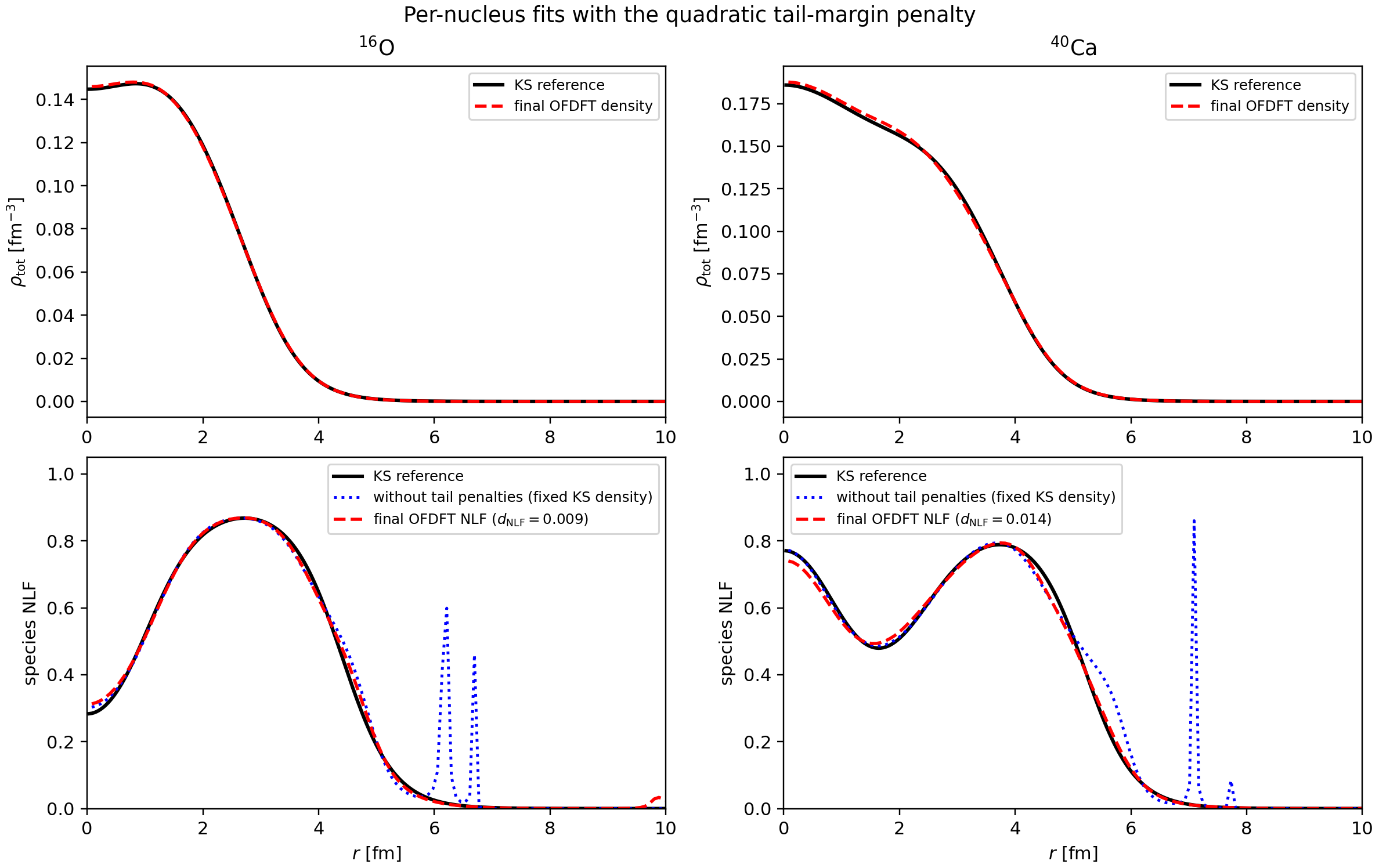}
\caption{Final per-nucleus functionals for $^{16}$O and $^{40}$Ca with
EL-potential matching and the quadratic tail-margin penalty. Top: final total
nucleon densities versus KS. Bottom: species-resolved NLF at the final ITE
density; the blue dotted NLF curves omit the tail penalty and contain the
artificial peak caused by a $\chi$ zero crossing near $r=6.7$~fm.}
\label{fig:light}
\end{figure}

\begin{figure}[tb]
\includegraphics[width=\linewidth]{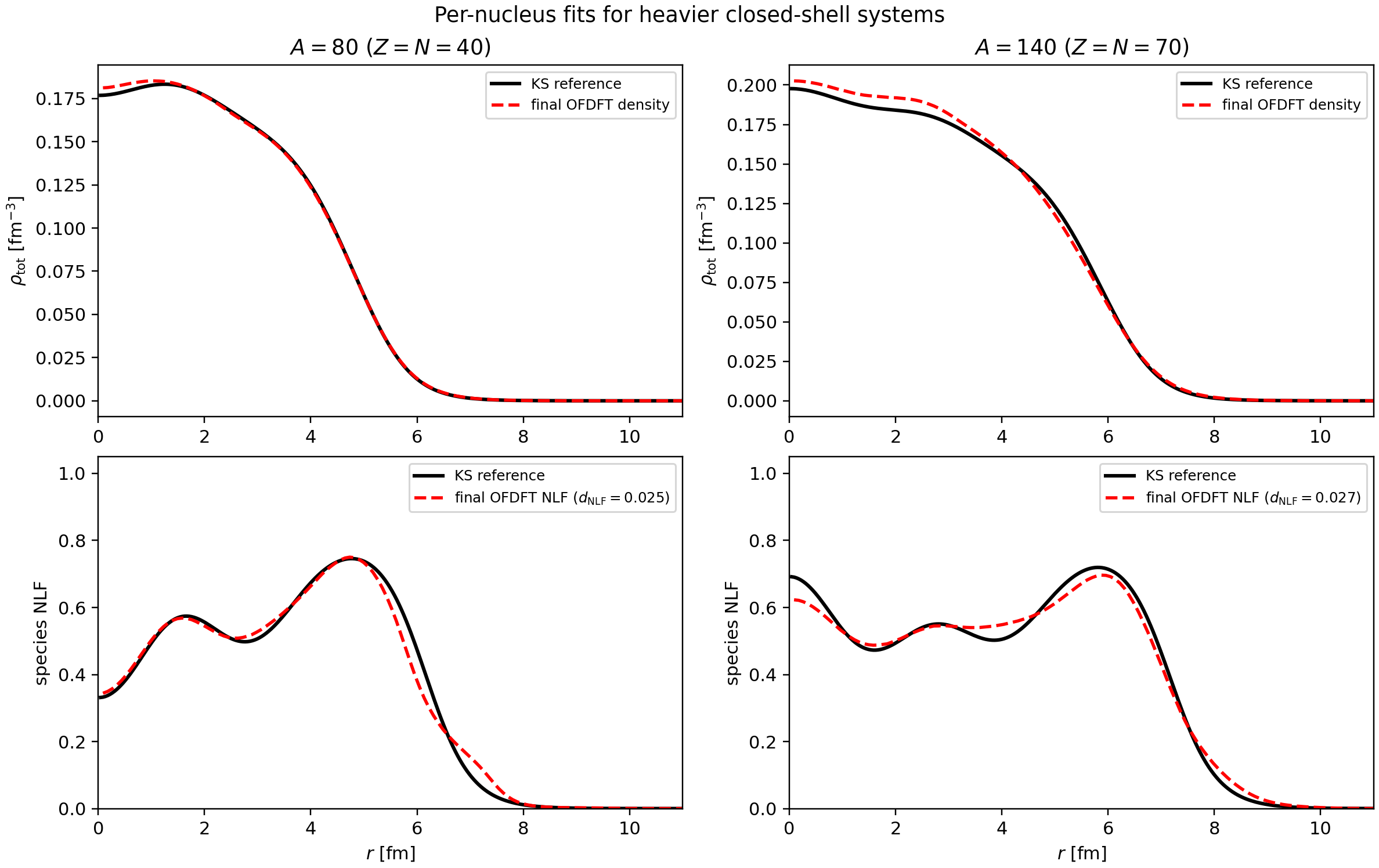}
\caption{Application of the per-nucleus procedure to the next two doubly
magic systems of the benchmark without spin--orbit coupling, $A=80$
($Z=N=40$) and $A=140$ ($Z=N=70$). Top: total nucleon density. Bottom:
species-resolved NLF (identical for neutrons and protons), including its three
to four interior extrema.}
\label{fig:heavy}
\end{figure}

\begin{table*}[tb]
\caption{Per-nucleus learned functionals: final adaptive-step ITE densities
versus the KS reference. $E_{\rm KS}$ is the independent KS total energy
obtained with the same simplified SkP interaction and the same mass constant
as the released solver ($\hbar^2/2m=20.73553\,\mathrm{MeV\,fm^2}$;
regenerated by the released script \texttt{ks\_total\_energy.py}).
$E_{\rm OF}^{(64)}\equiv E_{\rm OF}^{(64)}[\rho_{\rm ITE}^{(24)}]$ is the
OFDFT energy reevaluated with $n_\mu=64$ on the final species density produced
by the $n_\mu=24$ ITE trajectory. The magnitude
$|\Delta E_{\rm ITE}^{(24)}|=|E_{\rm end}^{(24)}-E_{1}^{(24)}|$ uses the first
and final monitored energies of that trajectory; accepted monitor energies
are non-increasing. $R_{{\rm EL},q}^{(24)}$ is the final unnormalized
specieswise residual of Eq.~(\ref{eq:elresid}). The distance $d$ is defined by
Eq.~(\ref{eq:nlfdistance}). In entries $x\,(x_{\rm KS})$, the first value is
the final OFDFT result and the parenthesized value is the KS reference. Radii
are in fm, energies and residuals in MeV, and central total densities in
fm$^{-3}$.}
\label{tab:pernucleus}
\setlength{\tabcolsep}{8pt}
\begin{ruledtabular}
\begin{tabular}{lccccccc}
 & $E_{\rm KS}$ & $E_{\rm OF}^{(64)}$ & $|\Delta E_{\rm ITE}^{(24)}|$ & $R_{{\rm EL},q}^{(24)}$ [MeV] & $\rho_c$ (KS) & $r_{\rm rms}$ (KS) & $d$ \\
\hline
$^{16}$O  & $-126.50$ & $-126.89$ & $0.02$ & 0.017 & 0.146 (0.145) & 2.741 (2.741) & 0.009 \\
$^{40}$Ca & $-398.89$ & $-400.54$ & $0.34$ & 0.054 & 0.188 (0.186) & 3.402 (3.396) & 0.014 \\
$A{=}80$  & $-895.74$ & $-898.14$ & $0.20$ & 0.045 & 0.181 (0.177) & 4.104 (4.092) & 0.025 \\
$A{=}140$ & $-1677.4$ & $-1683.4$ & $4.1$  & 0.049 & 0.203 (0.198) & 4.811 (4.807) & 0.027 \\
\end{tabular}
\end{ruledtabular}
\end{table*}

Figure~\ref{fig:light} and Table~\ref{tab:pernucleus} summarize the final
per-nucleus functionals. The reevaluated endpoint energies lie below the KS
values by $0.4$--$6.0\MeV$ ($0.3$--$0.4\%$): the exact energy equalities are
imposed at the fixed KS densities, and the subsequent self-consistent
relaxation lowers the energy of the approximate functional below that
constrained point. For $^{16}$O the adaptive-step ITE density reproduces
the KS central NLF dip ($0.28\to0.31$), the surface peak ($0.87\to0.87$), and
the tail decay to below ${\rm NLF}=10^{-3}$ at $r=7.5$~fm; the weighted NLF
distance is $d=0.009$. For $^{40}$Ca the double-hump structure is reproduced
($0.48/0.50$ at the interior dip) with $d=0.014$. Two tail-specific
ingredients are required: (i)~the tail band, which suppresses an artificial oscillation of the
learned $\chi$ around the exponentially growing $\chi_{\rm KS}$, and (ii)~the
one-sided tail penalty, which removes artificial ${\rm NLF}=1$ peaks caused by $\chi$ zero
crossings; both artifacts live at $\rho\lesssim 10^{-6}\,{\rm fm}^{-3}$ and
are invisible to density-weighted value fits.

Applying the same fitting and relaxation procedure to the next two
oscillator-magic systems
(Fig.~\ref{fig:heavy}) shows that the procedure remains numerically stable
across the systems tested: the richer shell
structure (three to four interior NLF extrema) is tracked across the interior
region; radii agree to $0.004$--$0.012$~fm, and the central densities agree
to $2.5\%$ for both systems. The remaining systematic degradation is also
quantified: the monitored energy drift over the fixed 8000-step budget grows
to $4.1\MeV$ for $A=140$ ($0.24\%$ of $E$), while it stays at or below
$0.34\MeV$ for the three lighter systems.

\section{Results II: shared multi-nucleus fit and test on an unseen nucleus}
\label{sec:results2}

\begin{figure*}[tb]
\includegraphics[width=0.95\textwidth]{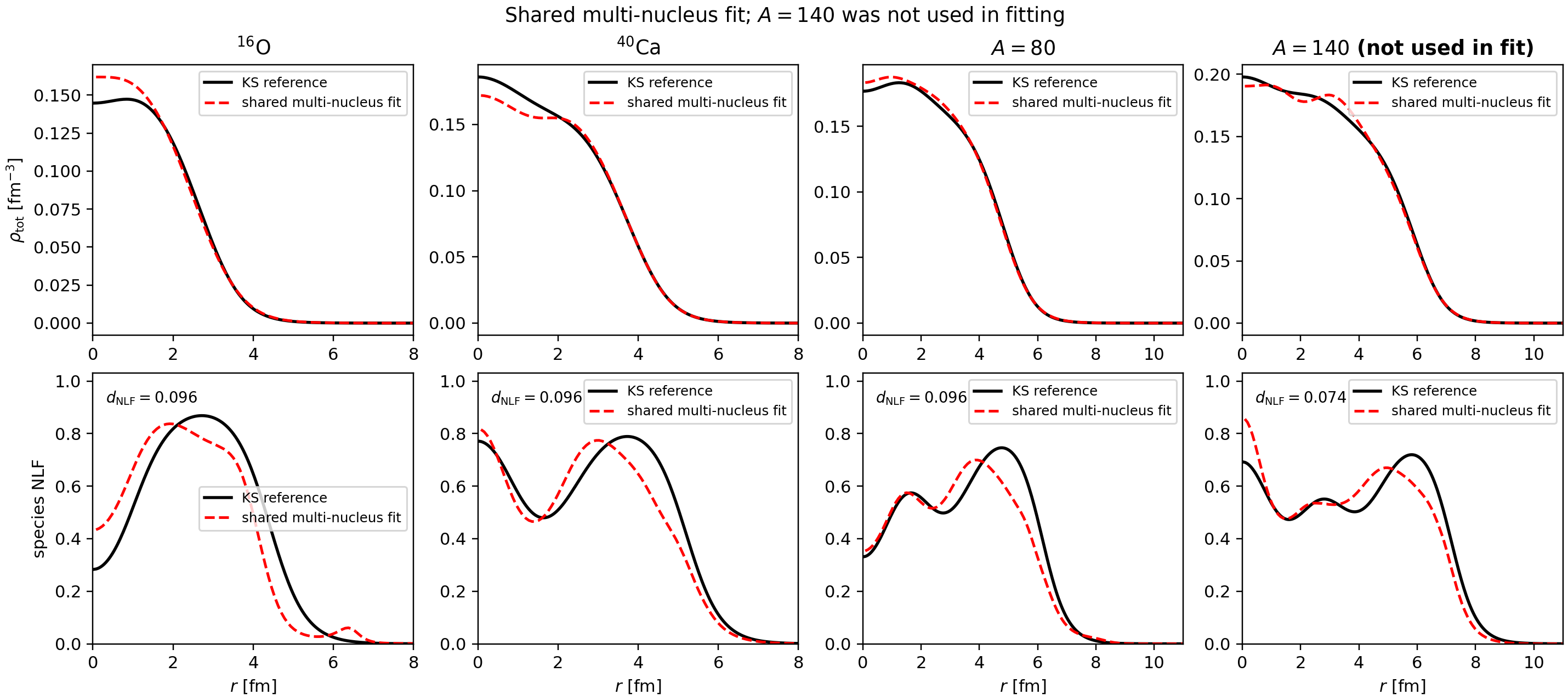}
\caption{Shared multi-nucleus (``universal'') functional: a \emph{single} kernel correction
$\delta g$ and local-correction shift, trained jointly on $^{16}$O, $^{40}$Ca and
$A=80$ with selected-path energy-rise penalties (built from the recorded
central-depletion direction of a preliminary fit; text) and per-nucleus
normalization of the tail-data weights; the production functional uses
$s_2=0$, i.e.\ no gradient regularization. Top: total nucleon density; bottom: species-resolved
NLF. None of the four final densities exhibits the identified central-depletion
or tail artifacts. The rightmost
column, $A=140$, is a holdout test: no $A=140$ observable or diagnostic enters either fitting
or model selection.}
\label{fig:universal}
\end{figure*}

\begin{table*}[tb]
\caption{Shared multi-nucleus functional (single $\delta g$ and shift; trained on
$^{16}$O, $^{40}$Ca, $A=80$; no $A=140$ observable or diagnostic enters
fitting or model selection). Same conventions as
Table~\ref{tab:pernucleus}. $E_{\rm OF}^{(64)}$ is
$E_{\rm OF}^{(64)}[\rho_{\rm ITE}^{(24)}]$, evaluated at $n_\mu=64$ on the
final $n_\mu=24$ ITE density; the production shared functional uses
$s_2=0$, so no gradient-regularization energy enters.
$\Delta E_{\rm KS}=E_{\rm OF}^{(64)}-E_{\rm KS}$. The labels ``tr.'' and ``h.o.''
denote a training nucleus and the holdout nucleus, respectively. The last
column $d$ is the density-weighted NLF distance to the KS reference, as in
Table~\ref{tab:pernucleus}; the parenthesized entries in the density and radius
columns are again the KS reference values.
The exact energy equalities hold at the fixed KS training densities; the
subsequent self-consistent relaxation lowers each training energy by
$1.3$--$3.2\MeV$ below $E_{\rm KS}$.}
\label{tab:universal}

\setlength{\tabcolsep}{10pt}
\begin{ruledtabular}
\begin{tabular}{llccccc}
 & & $E_{\rm OF}^{(64)}$ & $\Delta E_{\rm KS}$ & $\rho_c$ (KS) & $r_{\rm rms}$ (KS) & $d$ \\
\hline
$^{16}$O  & tr. & $-127.84$ & $-1.3$ & 0.162 (0.145) & 2.774 (2.741) & 0.096 \\
$^{40}$Ca & tr. & $-401.13$ & $-2.2$ & 0.172 (0.186) & 3.404 (3.396) & 0.097 \\
$A{=}80$  & tr. & $-898.91$ & $-3.2$ & 0.183 (0.177) & 4.088 (4.092) & 0.096 \\
$A{=}140$ & \textbf{h.o.} & $-1719.08$ & $-41.7$ & 0.190 (0.198) & 4.793 (4.807) & 0.074 \\
\end{tabular}
\end{ruledtabular}
\end{table*}

To test transfer across nuclei we train a \emph{single} $(\delta g,s)$ on
$^{16}$O$+{}^{40}$Ca$+A{=}80$ (per-nucleus data groups, three exact energy
equalities, and per-nucleus chemical potentials as auxiliary parameters).
All choices of functional form, basis size, relative data weights, hyperparameters,
and selected-path energy-rise penalties were made using only these three training nuclei.
No $A=140$ observable or diagnostic was inspected during model selection.
Only after the model and its coefficients had been fixed was $A=140$
evaluated; it is therefore a holdout, meaning a previously unseen test
nucleus. Using only the training nuclei,
the fit design exposed a central-depletion failure whose diagnosis is itself
instructive and whose resolution motivated the selected-path energy-rise rows
of Sec.~\ref{sec:method}; the preliminary no-path fit and all diagnostics
below are released and regenerable
(\texttt{kernel\_fit.py shared\_nopath};
\texttt{results/preliminary\_correction/}). (i)~A preliminary
shared fit \emph{without} the energy-rise rows produced a qualitative failure
invisible at fit level (its fixed-density NLF distance for $A=80$, $0.082$,
is comparable to the released shared fit's fixed-density value): at the production
quadrature ($n_\mu=24$) the KS-initialized $A=80$ calculation collapses to an
unphysical central depletion ($\rho_c=0.061$ versus $0.177$).
(ii)~Neither the gradient term of Eq.~(\ref{eq:gradct}) (scanned up to
$s_2=2\,\mathrm{fm}^4$: $\rho_c=0.061$--$0.063$) nor a finer angular
quadrature ($n_\mu=48$: $\rho_c=0.063$) removes the depletion, which is
consistent with an unfavorable-curvature direction of the fitted correction
itself that is not controlled by the first-order fit; the base functional is
stable along the released trial paths (its path-energy changes are $+1.3$ to
$+10.2\MeV$).
(iii)~The depletion is a genuine, algorithm-independent fixed point: relaxed
from the KS density at the production setting ($n_\mu=24$, $s_2=0$), the
rearranged diagonalization converges (residual below $10^{-8}$) to a
centrally depleted density, and the 8000-step imaginary-time endpoint agrees
with it to a relative $L^2$ density difference of $2.0\times10^{-4}$, with
identical central density ($\rho_c=0.061$ for both). Including the three energy-rise rows built from this recorded
artifact direction (Sec.~\ref{sec:method}; the endpoint density and the
trial densities are released in \texttt{scripts/path\_data/}) removes the failure at its source: the released
shared functional keeps the path energies near zero
($\Delta E=+0.41$, $+0.29$, $+1.09\MeV$ at the three released trial
densities---all uphill, though below the $2\MeV$ soft target), and its
KS-initialized $A=80$ relaxation is a stable, centrally filled profile already
at $s_2=0$ ($\rho_c=0.183$ versus the KS value $0.177$;
Table~\ref{tab:universal}).

At the working $n_\mu=24$ discretization the artifact is a localized central
depletion ($r\lesssim2$~fm, where the $2s$ orbital contributes), while the
surface is nearly unchanged. It can be regularized by supplementing the
nonlocal kernel with a single \emph{gradient regularization term}, added to the
local correction of Eq.~(\ref{eq:tau}),
\begin{equation}
E_{\nabla}[\rho_n,\rho_p] = s_2\,\frac{\hbar^2}{2m}\sum_{q=n,p}\int
d^3r\;\rho_q^{1/3}\,|\nabla\rho_q|^2 ,
\label{eq:gradct}
\end{equation}
with a single coefficient $s_2$ of dimension $\mathrm{fm}^4$. The dimensions
follow from $\rho_q^{1/3}|\nabla\rho_q|^2d^3r\sim\mathrm{fm}^{-6}$ and
$C\sim\mathrm{MeV\,fm^2}$, so $s_2C$ converts the integral to energy. Its
continuum functional derivative is
\begin{equation}
v_{\nabla,q} = s_2\,\frac{\hbar^2}{2m}\left[\tfrac13\rho_q^{-2/3}|\nabla\rho_q|^2
 - 2\,\nabla\!\cdot\!\big(\rho_q^{1/3}\nabla\rho_q\big)\right],
\label{eq:gradv}
\end{equation}
where surface terms vanish for the stated Dirichlet/tail boundary conditions.
The first term comes from differentiating $\rho_q^{1/3}$ and the divergence
term from integrating the variation of $\nabla\rho_q$ by parts.
The potential $v_{\nabla,q}$ is included in the full kinetic potential $v_{T,q}$, the
rearranged auxiliary potential $v_{{\rm rest},q}$, and the ITE local potential
$v_{\rm loc}$. In the code the potential is not this
continuum expression but the exact discrete adjoint of the first-difference
energy (Appendix~\ref{app:numerics}), so it matches finite differences of the
energy to round-off. The $\rho_q^{1/3}$ prefactor makes the term
vanish for uniform matter---so it is uniform-density-neutral and does not
change the uniform-matter (Thomas--Fermi) limit---and
removes the low-density singularity of a plain Weizs\"acker-like form, so its
Euler--Lagrange potential stays finite in the exponential tail. It does not,
however, repair the small uniform-limit error of the finite-table kernel
itself. It penalizes spatial density gradients generally; for the tested
$A=80$ trajectories its largest numerical effect is in the central region.
For the preliminary correction the released $s_2$ scan
(\texttt{results/preliminary\_correction/}) shows that the gradient term
does \emph{not} remove the depletion: over the 8000-step budget the
KS-initialized central density stays at $\rho_c=0.061$--$0.063$ for
$s_2=0$--$2\,\mathrm{fm}^4$, and the depletion equally persists at the finer
$n_\mu=48$ quadrature. The instability is thus a property of the no-path fit
itself, not of the discretization, and it is removed only at the fit stage by
the energy-rise rows. The released shared functional needs no gradient
regularization either, and adding one would not improve it: a targeted spot
check at the production setting moves the KS-initialized $A=80$ central
density from $\rho_c=0.1834$ at $s_2=0$ to $0.1988$ at
$s_2=0.5\,\mathrm{fm}^4$, i.e.\ further from the KS value $0.1769$ rather
than closer to it. The production shared functional therefore uses $s_2=0$;
the gradient term of Eq.~(\ref{eq:gradct}) remains implemented and released
as optional machinery, with its exact discrete adjoint retained in the
solver.

Figure~\ref{fig:universal} and Table~\ref{tab:universal} show the released
shared functional ($s_2=0$). All four final densities
are free of the artificial tail peaks, central depletion, and oscillations
described above. The transfer properties separate cleanly:
\begin{itemize}
\item \emph{NLF shell structure transfers.} The holdout distance $d=0.074$ is
comparable to the training values ($0.096$--$0.097$)---indeed the smallest of
the four; the interior
oscillations of the unseen $A=140$ system are reproduced across the interior
region. At $r=(1.5,3.0,6.0)$~fm, respectively, the KS values are
$(0.475,0.547,0.713)$ and the values at the self-consistent OFDFT density are
$(0.476,0.530,0.591)$, consistent with
structure-level transfer in this holdout test.
\item \emph{Radii transfer}: the universal-functional radii agree with KS to
$0.004$--$0.033$~fm across the four nuclei ($0.014$~fm for the $A=140$
holdout). Central densities are reproduced less accurately
($4$--$12\%$), with $^{16}$O at the top of that range; for the
released functional the $A=80$ central density shifts by a further $7.8\%$
between $n_\mu=24$ and $48$ (Appendix~\ref{app:numerics}).
\item \emph{Absolute energies do not transfer} at this basis size: the
holdout energy is overbound by $41.7\MeV$ ($2.5\%$) in the $n_\mu=64$ evaluation. During training the
per-nucleus energies are matched exactly on the fixed KS densities, but the
$A$~dependence of the energy integral is not captured by the shared 15 fit
parameters (14 kernel-basis coefficients and the shift $s$).
\end{itemize}
The working-discretization behavior above gives a narrower numerical lesson.
First-order supervision on reference densities does not by itself guarantee
stable relaxation after the functional and its quadrature have both been
discretized. Stability tests must therefore vary not only the initial density
and density-update algorithm but also the numerical representation of the nonlocal
kernel. In the present case the gradient term is an effective regularizer at
$n_\mu=24$, whereas the disappearance of the depleted trajectory at
$n_\mu=48$ prevents us from identifying a corresponding instability of the
quadrature-converged functional.

\section{Discussion}
\label{sec:discussion}

\begin{figure}[tb]
\includegraphics[width=\linewidth]{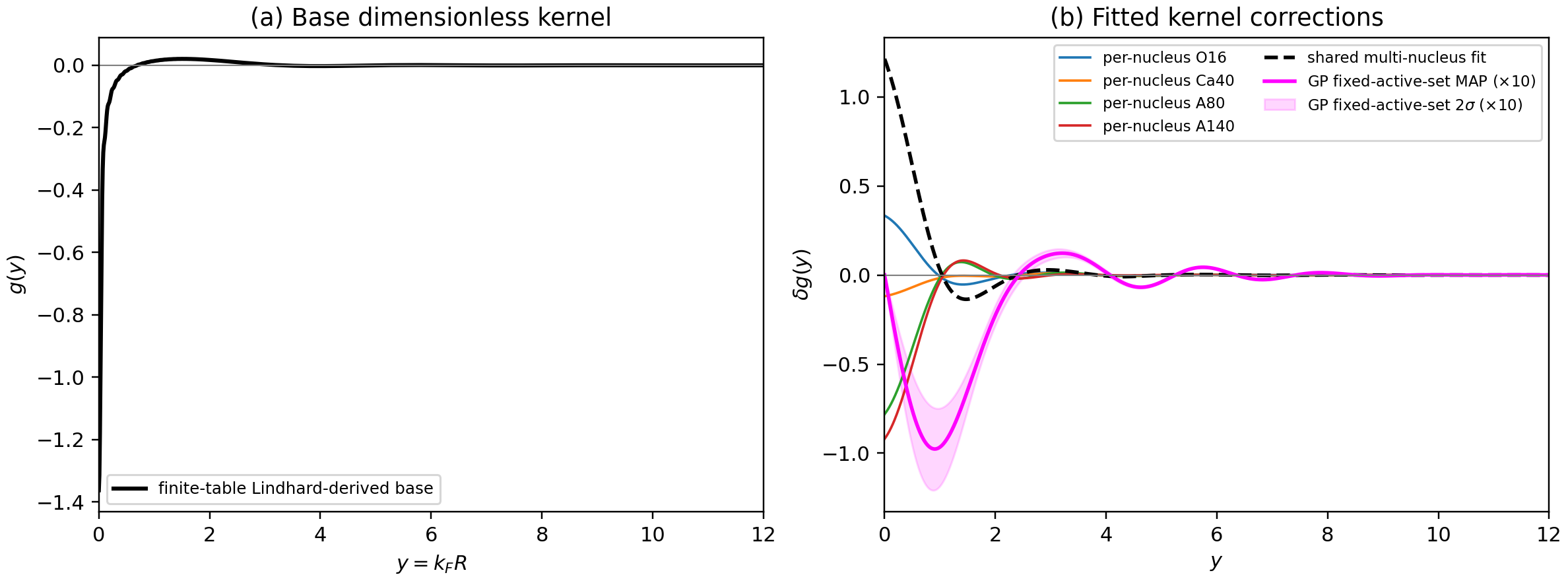}
\caption{(a)~Base dimensionless kernel $g(y)$. (b)~Learned corrections
$\delta g(y)$: per-nucleus fits, the shared multi-nucleus fit, and the
Gaussian-process posterior mean ($\times10$) with its fixed-active-set
$2\sigma$ band.}
\label{fig:shapes}
\end{figure}

\begin{figure}[tb]
\includegraphics[width=\linewidth]{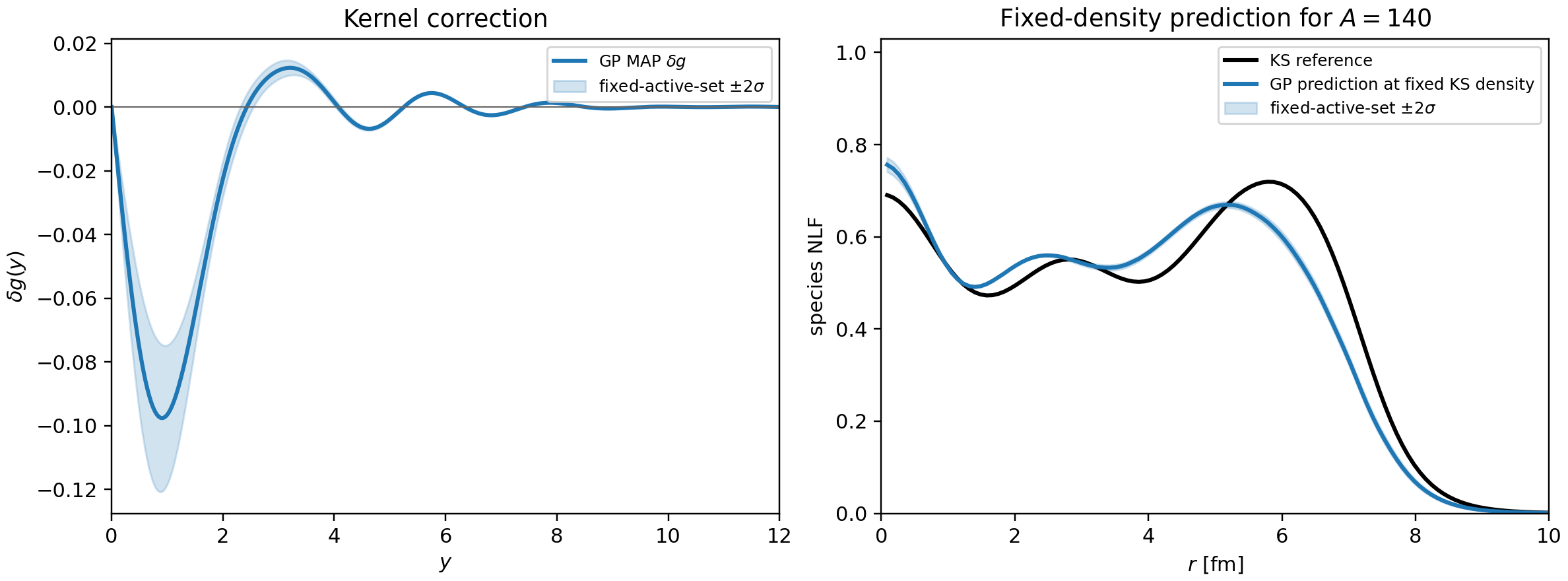}
\caption{Gaussian-process reformulation: posterior mean and pointwise
fixed-active-set $2\sigma$ band of $\delta g(y)$ (left); fixed-density NLF
prediction for the previously unseen
$A=140$, obtained by propagating the posterior coefficient covariance to
$\chi$ at $\rho_{\rm KS}$ and then to the NLF (right). Here $\sigma$ is the
local inverse-Hessian posterior standard deviation with the selected
hyperparameters and final active penalty rows held fixed.}
\label{fig:gp}
\end{figure}

\emph{What is learned.} Figure~\ref{fig:shapes} compares the learned kernel
corrections. The per-nucleus $\delta g$ are dominated by structure at
$y\lesssim 3$, i.e., at inter-nucleon distances $R\lesssim 3/k_F$,
sharpening the short-range part of the kernel relative to the Lindhard-based
$g(y)$, with smaller adjustments in the first oscillation of the long-range
response generated by the sharp Fermi surface. The
shared correction has the same qualitative pattern. The Gaussian-process posterior mean
yields comparable fixed-density NLF accuracy with a much smaller peak
amplitude, $\max|\delta g|=0.098$ versus $4.5$ for the Gaussian-basis fit; its
fitted local-correction shift is $s=-8.32492$. This peak-amplitude comparison is not
a full effect-size statement---the two corrections carry different uniform
moments and pair with different local-correction shifts---but the selected smooth
prior shows that the large, largely canceling basis components are not
required to describe the training observations.

\emph{Statistical interpretation and uncertainty.} The Gaussian data groups
are linear functionals of $\delta g$ and therefore define GP regression
with operator observations (Appendix~\ref{app:gp}). The correlation length,
prior amplitude, and noise scale are selected by marginal likelihood on the
finite grid of Appendix~\ref{app:gp}; the reported point is not an
unconstrained evidence optimum. The soft inequality penalties are included at
the MAP level. The reported covariance is the inverse-Hessian Gaussian
approximation for the final quadratic subproblem, with the selected
hyperparameters and active penalty rows held fixed; it is not a posterior
conditioned on exact inequality constraints. It propagates coefficient uncertainty through the
linear response-ratio prediction at fixed $\rho_{\rm KS}$; it does not
propagate uncertainty through the nonlinear ITE relaxation and does not
include model discrepancy. On the $80$ radial points satisfying
$1<r<8$~fm and $\rho_q>10^{-5}\,\mathrm{fm}^{-3}$, the conditional $2\sigma$
band covers $10/80=0.125$ of the $A=140$ response-ratio residuals. The
remaining discrepancy is consistent with limited model expressivity, though
likelihood/noise misspecification, hyperparameter uncertainty, and the
fixed-density linear prediction are also possible contributors.

\emph{Limitations and outlook.} Four limitations are quantified above: the
non-transfer of absolute energies ($-41.7\MeV$ at $A=140$), the residual
smooth NLF undershoot of the shared fit in the $^{16}$O surface
($0.20$ versus $0.41$ at $r=4.5$~fm, a shared-capacity effect), and the
growth of the energy drift with mass in per-nucleus fits, together with the
residual percent-level quadrature sensitivity of the $A=80$ central
density. The first three motivate enrichment of the kernel-correction
space (including possible
$t$-dependence of $\delta g$ beyond the shape correction), more training
systems, and explicit uniform-matter constraints anchoring the
$A$~dependence of the energy. The fourth requires higher-order or more
efficient angular integration before the $A=80$ central density or the
$n_\mu=24$ depletion artifact can be interpreted as a property of the
continuum functional. The present gradient-dependent term should therefore
be viewed as a limit-preserving numerical regularizer for the production
discretization, not as a uniquely inferred physical correction. Extension toward
realistic functionals requires spin-orbit and Coulomb terms, isospin
asymmetry, and deformation. The linear-in-kernel fitting strategy can be
retained, but such extensions require suitable density-level representations
of the additional terms. The benchmark setting adopted here is that of Ref.~\cite{WuPRL2026},
which allows direct comparison and isolates the shell-structure question.

\section{Summary}
\label{sec:summary}

We have demonstrated self-consistent orbital-free nuclear DFT calculations
with a learned nonlocal kinetic energy density functional. The key technical
elements are (i)~an analytic derivative of the represented finite-table
density-dependent-kernel nonlocal term; (ii)~functional-derivative-informed (Sobolev)
fitting that exploits the linearity of the functional in the kernel shape;
and (iii)~explicit stability checks across density-update algorithms, initial
conditions, and numerical quadratures. A central-depletion instability exposed by a preliminary
shared fit persists under gradient regularization and under a finer angular
quadrature, and is removed only at the fit stage by the selected-path
energy-rise penalty; a single limit-preserving gradient term is implemented
and released but not required by the final functional. The per-nucleus fits impose exact KS-energy
matching at the fixed KS training densities; their final self-consistent
energies are reported separately. They track the density and the full
NLF shell structure of doubly-magic $N=Z$ systems from $A=16$ to $A=140$,
with the remaining central-density deviations quantified; a single universal
correction trained on $A\le 80$ transfers the shell structure and radius to a
previously unseen $A=140$ test system, while absolute
energies remain nucleus-specific at the present basis size. For the base
constant-$k_F$ $^{16}$O benchmark, the rearranged diagonalization converges
without mixing in 41 density-update iterations at $\lambda=3$ and agrees with a
prolonged ITE trajectory to $0.368$~keV and a relative density difference of
$4.59\times10^{-4}$. The learned-functional observables reported here are
evaluated on the final ITE densities, which are cross-checked by the
rearranged diagonalization. A Gaussian-process reformulation supplies
finite-grid marginal-likelihood selection and conditional parameter-uncertainty
quantification. These results
turn the nonlocal-functional
program of Ref.~\cite{WuPRL2026} into a self-consistent scheme and identify
the concrete ingredients---analytic derivatives, solver cross-checks,
quadrature control, and energy anchoring---needed for a universal orbital-free
nuclear functional.

\section*{Data availability}
The solver, all calculation inputs, the analysis scripts, and the data
underlying the tables and figures are available at
\url{https://github.com/fimoto/nofdft-nuclear-kedf}.

\begin{acknowledgments}
The author received no funding or other support and acknowledges the use of AI-based assistants---Claude (Claude Opus
4.8, Anthropic) and GPT-5.6 (OpenAI)---as interactive aids to develop and debug some part of the author's numerical codes during this work.
\end{acknowledgments}

\appendix

\section{Imaginary-time-evolution details}
\label{app:solver}

\emph{ITE scheme.} Define the real density amplitude
$\phi_q(\bm r)=\sqrt{\rho_q(\bm r)}$, with units fm$^{-3/2}$ and
$\int d^3r\,|\phi_q|^2=N_q$. Suppressing $q$, one symmetric second-order
split step for the Hamiltonian frozen at iteration $k$ is
\begin{equation}
 \phi^{(k+1)}={\cal P}_{N_q}
 e^{-\Delta t v_{\rm loc}[\rho^{(k)}]/2}
 e^{-\Delta t\hat T_{\rm vW}}
 e^{-\Delta t v_{\rm loc}[\rho^{(k)}]/2}\phi^{(k)}.
\label{eq:itesplit}
\end{equation}
Here $\Delta t$ is an imaginary-time step with units MeV$^{-1}$,
${\cal P}_{N_q}$ rescales the updated amplitude to particle number $N_q$,
and $\hat T_{\rm vW}=-C\nabla^2$ is the von~Weizs\"acker kinetic operator.
The multiplicative remainder is
$v_{\rm loc}=v_{\rm kernel}+v_{\rm corr}+v_{\nabla}+v_{\rm SkP}$, where
$v_{\rm kernel}\equiv v_{\rm NL}$ includes the learned kernel correction,
$v_{\rm corr}$ is the response of the local Thomas--Fermi correction,
$v_{\rm SkP}=\delta E_{\rm int}^{\rm SkP}/\delta\rho_{\rm tot}$, and
$v_{\nabla}=0$ for $s_2=0$. The
von~Weizs\"acker propagator is applied to $u=r\phi$ with the Dirichlet
second-difference operator at the $n_r=250$ interior points
$r_i=i\Delta r$, where $\Delta r=R_{\rm box}/(n_r+1)$ and
$R_{\rm box}=20$--$22$~fm. The local exponential argument is clamped to
$[-80,80]$ to prevent overflow. The fixed-step split update itself has no energy-based rejection; only the
adaptive controller (below) rejects and rolls back. Energies are monitored at
the first step and then every $20$ iterations for the per-nucleus runs and
every $50$ for the shared multi-nucleus runs. We distinguish the number of \emph{attempted}
steps (the iteration budget, here $8000$) from the number of \emph{accepted}
monitor points (the scheduled monitor points that pass the energy-acceptance
test); with the adaptive controller the
monitored energy is non-increasing across the accepted points by construction,
whereas for a plain fixed-step run monotonic decrease is only an observed
property. The final summary is recomputed on the written final density.

The explicit-potential step has a stiffness-dependent step-size ceiling
($\Delta t\lesssim5\times10^{-5}\,\mathrm{MeV}^{-1}$; the von~Weizs\"acker
propagator itself is unconditionally stable, being an exact exponential of
the frozen second-difference operator, so the ceiling comes from the explicit
local potential and the splitting error, not from the kinetic term). A small
fixed step then relaxes the light nuclei fully but leaves the heavy systems
short of stationarity: at $\Delta t=10^{-5}$ over $8000$ steps the final EL
residuals $\|\hat h\phi-\mu\phi\|_{L^2}$ are $0.03$/$0.09$
($^{16}$O/$^{40}$Ca) but $0.74$/$0.52$ ($A{=}80$/$A{=}140$) per-nucleus, the
heavy densities being geometrically settled ($\rho_c$, $r_{\rm rms}$ stable to
$<1\%$) but the functional not yet stationary.
The reported species residual is $R_{{\rm EL},q}$ of
Eq.~(\ref{eq:elresid}); its normalization and units therefore remain the same
in the ITE and diagonalization checks. We adapt the step size as follows. At
an accepted monitor point, $\Delta t$ is multiplied by $1.1$, up to
$5\times10^{-5}\,\mathrm{MeV}^{-1}$. If the monitored energy exceeds the
previous accepted value by more than $10^{-12}\MeV$, the trial density is
rejected, the previous accepted density is restored, and $\Delta t$ is
multiplied by $0.5$, with a lower bound of
$10^{-8}\,\mathrm{MeV}^{-1}$. This drives the heavy per-nucleus residuals to
$0.02$--$0.08$ at the same $8000$-step budget without changing the converged
geometry, and yields the converged densities quoted in
Tables~\ref{tab:pernucleus}--\ref{tab:universal}. As an independent check, the
rearranged diagonalization reaches the same fixed points (Sec.~\ref{sec:solver}).
The symmetric split form is held fixed throughout the comparisons.

\emph{Initialization.} Production runs are initialized at the KS density
(Sec.~\ref{sec:solver}). As a global-convergence probe we also evolved the
$^{16}$O per-nucleus functional from the smooth, featureless profile
\begin{align}
 \rho_q(r)&=\frac{\rho_{0q}}{1+\exp[(r-r_0)/a_F]},\nonumber\\
 4\pi\int_0^{R_{\rm box}}r^2\rho_q(r)\,dr&=N_q,
\end{align}
with $r_0=3.0$~fm; the normalization determines $\rho_{0q}$ and
$a_F=0.55$~fm is the surface diffuseness. The monotone flow closes $\sim98\%$ of
the energy gap over $4.0\times10^{4}$ iterations
($E=-124.6$ versus $-126.9\MeV$; final EL residual $4.2$), with the density
approaching the KS-initialized final density (weighted $L^2$ difference $8\%$;
$r_{\rm rms}=2.89$ versus $2.74$~fm), with no tendency toward a distinct
self-consistent solution along the observed trajectory. Further relaxation from featureless
profiles is limited by slow imaginary-time dynamics far from the
KS-initialized final density at the fixed small step. This is a solver-efficiency
issue, not evidence for a separate minimum of the learned functional.

\section{Numerical verification and settings}
\label{app:numerics}

Analytic EL potentials are verified against second-order central finite
differences of the total energy (relative agreement down to
$\sim10^{-11}$ at the optimal finite-difference perturbation amplitude), separately for the base kernel and for
each $\delta g$ basis function. The gradient regularization term of
Eq.~(\ref{eq:gradct}) is implemented so that its analytic potential is the
\emph{exact discrete adjoint} of the first-difference energy. For one species,
\begin{equation}
 E_{\nabla,q}^{\rm disc}
 =s_2C\sum_i W_i\rho_{q i}^{1/3}(D\boldsymbol\rho_q)_i^2,
 \qquad E_\nabla^{\rm disc}=\sum_{q=n,p}E_{\nabla,q}^{\rm disc},
\label{eq:graddisc}
\end{equation}
where $\boldsymbol\rho_q=(\rho_{q1},\ldots,\rho_{q n_r})^{\rm T}$,
$W_i=4\pi r_i^2w_i$ contains the implemented radial quadrature weight
$w_i$, and $D$ uses centered differences in the interior and one-sided
differences at the two endpoints. With
$W=\operatorname{diag}(W_1,\ldots,W_{n_r})$ and $\odot$ denoting
elementwise multiplication, the potential placed on the same grid is
\begin{align}
 \boldsymbol v_{\nabla,q}^{\rm disc}={}&s_2C\left[
 \frac{1}{3}\boldsymbol\rho_q^{-2/3}\odot
 (D\boldsymbol\rho_q)^2\right.\nonumber\\
 &\left.+2W^{-1}D^{\rm T}W
 \{\boldsymbol\rho_q^{1/3}\odot D\boldsymbol\rho_q\}
 \right].
\label{eq:gradadjoint}
\end{align}
Thus $v_{\nabla,q,i}^{\rm disc}=W_i^{-1}
\partial E_{\nabla,q}^{\rm disc}/\partial\rho_{q i}$. Density powers are
evaluated with the code floor $\max(\rho_{q i},10^{-20}\,\mathrm{fm}^{-3})$
to avoid numerical singularities. The resulting directional derivative
matches finite differences to round-off
($\sim3\times10^{-11}$ relative), like the von~Weizs\"acker and Skyrme
terms, and the potential remains bounded in the exponential tail
($|v_\nabla|\lesssim10^{-3}$ for $r>8$~fm) because of the $\rho^{1/3}$
prefactor. The von~Weizs\"acker and Skyrme-gradient
terms use the same centered-interior/one-sided-endpoint first derivative
(with exact discrete adjoints in the corresponding gradients), while the imaginary-time
kinetic propagator uses the Dirichlet second-difference operator. On the
smooth final production densities, the two integrated von~Weizs\"acker energies
differ by $0.023$--$0.105\MeV$ per species ($0.04$--$0.06\%$). The ITE is
therefore not the exact discrete gradient flow of the first-difference
energy, although the integrated discrepancy is small on the reported smooth
densities. The linearity of
the functional in $g$ is verified by direct response-assembly tests, and an
independent high-order angular quadrature with $n_\mu=96$ checks the angular
convolution separately from the monitored production discretization.

Three angular settings are used for different stages of the calculation;
Table~\ref{tab:quadrature_roles} states both the density on which each is
evaluated and the quantity reported.
\begin{table}[b]
\caption{Roles of the angular quadratures. ``Final ITE(24) density'' means
the density produced by the $n_\mu=24$ trajectory; reevaluation does not
restart the SCF evolution.}
\label{tab:quadrature_roles}
\begin{ruledtabular}
\begin{tabular}{lccc}
Purpose & density & $n_\mu$ & reported quantity \\
\hline
ITE trajectory & evolving & 24 & density, $\Delta E$, $R_{\rm EL}$ \\
Energy reevaluation & final ITE(24) & 64 & tabulated energy \\
NLF reevaluation & final ITE(24) & 48 & NLF curve and $d_{\rm NLF}$ \\
\end{tabular}
\end{ruledtabular}
\end{table}
Writing the Fourier momentum as $q=k_F\bar q$, the dimensionless transform
is truncated at $\bar q_{\max}=80$ (code keyword \texttt{xqmax}) and evaluated
by composite Simpson quadrature with $N_{\bar q}=6000$ even subintervals
(\texttt{nq}). The kernel is stored at $N_y$ uniformly spaced nodes, including
both endpoints, on $0\le y\le y_{\max}$, and $g$ and $g'$ are evaluated from
the same natural cubic spline; both are set to zero beyond $y_{\max}$. The
adaptive tables use $(y_{\max},N_y)=(102.718,12840)$ for the light systems and
$(112.990,14124)$ for the heavier systems. The $n_\mu=48$ and $64$
calculations are diagnostic reevaluations rather than the angular quadrature
that generated the ITE trajectories. The NLF is defined from the kinetic
kernel using the convention of Eq.~(\ref{eq:tau}); the local gradient
regularization term of Eq.~(\ref{eq:gradct}) enters the energy and the EL potential
but is not folded into this kinetic-energy-density convention, so the reported NLF
curves are a kernel diagnostic and are unchanged by $s_2$ at fixed density. The angular-quadrature
error is not negligible at the derivative level: core relative errors are at
the few-percent level, and the offset between the monitored $n_\mu=24$ energy
and the $n_\mu=64$ readout reaches $14.8\MeV$ for the $A=80$ per-nucleus
functional ($2.1\MeV$ for the shared functional). A targeted 8000-step ITE recomputation of the \emph{released} shared
functional at $n_\mu=48$ (released with the data,
\texttt{results/preliminary\_correction/nmu48\_released/}) shifts the $A=80$
endpoint moderately: the central density moves from $0.1834$ to
$0.1977\,\mathrm{fm}^{-3}$ ($7.8\%$), the rms radius by $0.004$~fm, the NLF
distance from $0.096$ to $0.094$, and the $n_\mu=64$ energy readout of the
endpoint by $0.12\MeV$. For the preliminary no-path correction the central
depletion persists unchanged at $n_\mu=48$. Thus the radius and integrated
NLF diagnostic are comparatively stable, while the $A=80$ central density
retains a percent-level angular-quadrature sensitivity. Using $n_\mu=64$ for all tabulated energies provides a
common comparison convention, not a convergence certificate.

Fit hyperparameters: the EL-potential-data weight is $\beta=2$
($\beta_{A=80}=12$ in the joint fit); the surface--tail-data weight is
$\eta_T=0.3$ ($1.2$ for $^{16}$O in the final shared multi-nucleus fit); the
tail-margin and selected-path energy-rise weights are $12$ and $10$, with
target margins $\chi_{\min}=5$ and $\varepsilon=2\MeV$, respectively. The sets of
violated rows and their fixed-within-solve normalizations are updated
between least-squares solves until they no longer change. This is the
active-set repeated penalized least-squares iteration of
Eqs.~(\ref{eq:ineqpen}) and (\ref{eq:activesets}). The released GP
implementation performs $16$ updates; its active rows and coefficients cease
to change at update $14$ for the released data. All reported GP quantities
are generated from the final update. The value, EL-potential, and tail data
groups are normalized per nucleus by their
weighted signal power.

\section{Gaussian-process formulation}
\label{app:gp}

The kernel correction is expanded in a reduced-rank Gaussian-process basis
\cite{SolinSarkka2020},
$\delta g(y)=\sum_{j=1}^{M}w_j\varphi_j(y)$, where
\begin{equation}
 \varphi_j(y)=\sqrt{\frac{2}{L}}\sin\!\left(\frac{\pi j y}{L}\right),
 \qquad 0\le y\le L.
\label{eq:gpbasis}
\end{equation}
The code sets $\varphi_j(y)=0$ for $y>L$. The retained rank is $M=20$ and
the dimensionless interval length is $L=12$. The mode index is
$j=1,\ldots,M$, $w_j$ is its dimensionless random coefficient, and the basis
functions are orthonormal on $[0,L]$. Defining
$\Omega_j=\pi j/L$, the prior is
\begin{align}
 w_j&\sim{\cal N}[0,S_{\rm SE}(\Omega_j)],\nonumber\\
 S_{\rm SE}(\Omega)&=\sigma_f^2\sqrt{2\pi}\,\ell_{\rm GP}
 \exp(-\ell_{\rm GP}^2\Omega^2/2),\nonumber\\
 S_{\rm GP}&=\operatorname{diag}\{S_{\rm SE}(\Omega_1),\ldots,
 S_{\rm SE}(\Omega_M)\}.
\label{eq:gpprior}
\end{align}
Here $\ell_{\rm GP}$ is the correlation length in the dimensionless $y$
coordinate and $\sigma_f$ is the prior amplitude. The distinct symbol
$\ell_{\rm GP}$ avoids confusion with orbital angular momentum.

The full vector solved by the GP implementation is
\begin{equation}
 \bm\vartheta=(w_1,\ldots,w_M,s,\mu_1,\ldots,
 \mu_{N_{\rm tr}})^{\rm T},\qquad N_{\rm tr}=3,
\label{eq:gpvector}
\end{equation}
where $s$ is the local-correction shift and the last three entries are the
auxiliary chemical potentials of the training nuclei. The prior precision
$S_{\rm GP}^{-1}$ acts on the $w_j$ block. Weak fixed diagonal regularizers
$10^{-4}$ and $10^{-10}$ are applied to $s$ and the $\mu_\nu$ entries,
respectively, only to regularize otherwise uninformative directions. All
Gaussian data groups of Sec.~\ref{sec:method} are linear in
$\bm\vartheta$, and the MAP coefficients are obtained with the same bordered
KKT construction used for the exact energy equalities.

Hyperparameters $(\ell_{\rm GP},\sigma_f)$ and a global likelihood scale
$\kappa$ are compared by log marginal likelihood over the finite grid
$\ell_{\rm GP}\in\{0.6,1,1.5,2.2\}$, $\sigma_f\in\{0.3,1,3\}$, and
$\kappa\in\{30,100,300,1000\}$. The scalar $\kappa$ multiplies the common
precision of the normalized Gaussian observation groups; it changes their
likelihood weight relative to the GP prior without changing their internal
relative weights. The finite-grid maximum is
$(\ell_{\rm GP},\sigma_f,\kappa)=(1.5,0.3,300)$; it is not an unconstrained optimum,
and $\sigma_f$ lies at the grid edge. Fixing the noise scale to the signal
power drives the finite-grid evidence toward $\delta g\to0$, motivating the
separate likelihood scale. The evidence scores only the Gaussian value,
EL-potential, and surface--tail data groups. The exact energy equalities and
soft quadratic penalties enter only at the MAP stage. The released
implementation performs 16 active-set updates; the active rows and
coefficients cease to change at update 14 for the released data. The
resulting correction has
$\max|\delta g|=0.098$ and local-correction shift $s=-8.32492$.

For completeness, the released GP implementation specializes the weights in
Eq.~(\ref{eq:ineqpen}) as follows. If $\chi^{(0)}_{\nu i}$ denotes the
response ratio before the fitted GP correction, define
\begin{equation}
 \widetilde Z_{T,\nu}^{(k)}=
 \sum_{i\in{\cal A}_{T,\nu}^{(k)}}
 [\chi_{\min}-\chi^{(0)}_{\nu i}]^2 .
\end{equation}
The two normalization scales are
\begin{align}
 Z_{T,\nu}^{(k)}&=\max\{\widetilde Z_{T,\nu}^{(k)},10^{-12}\},\nonumber\\
 Z_S&=(10\MeV)^2.
\label{eq:gpnormalizers}
\end{align}
For compactness, define the active squared residuals
\begin{align}
 {\cal R}_{T,\nu}^{(k)}(\bm\vartheta)&=
 \sum_{i\in{\cal A}_{T,\nu}^{(k)}}
 [\chi_{\nu i}(\bm\vartheta)-\chi_{\min}]^2,\nonumber\\
 {\cal R}_{S}^{(k)}(\bm\vartheta)&=
 \sum_{(\nu,j)\in{\cal A}_{S}^{(k)}}
 [\Delta E_{\nu j}(\bm\vartheta)-\varepsilon]^2,\nonumber\\
 \overline{\cal R}_{T}^{(k)}(\bm\vartheta)&=
 \sum_\nu\frac{{\cal R}_{T,\nu}^{(k)}(\bm\vartheta)}
 {Z_{T,\nu}^{(k)}}.
\end{align}
The quadratic contribution held fixed during update $k+1$ is then
\begin{align}
 {\cal P}_{\rm ineq}^{(k)}(\bm\vartheta)={}&
 \frac{\kappa\eta_H}{2}\overline{\cal R}_{T}^{(k)}(\bm\vartheta)
 \nonumber\\
 &+\frac{\kappa\eta_S}{2Z_S}{\cal R}_{S}^{(k)}(\bm\vartheta).
\label{eq:gpineqweights}
\end{align}
Here $\eta_H=12$ and $\eta_S=10$ are the released tail and selected-path
penalty weights.
Equations~(\ref{eq:gpnormalizers}) and (\ref{eq:gpineqweights}) reproduce the
normal-matrix additions in the released \texttt{gp\_fit.py}. The selected
tail grid points enter with equal weight; there is no additional radial
quadrature weight in this penalty. The $10^{-12}$ floor only avoids division
by zero. $Z_{T,\nu}^{(k)}$ is recomputed when the tail active set changes and
then held fixed during the following equality-constrained least-squares
solve. By contrast, $Z_S$ is constant.

All three selected-path penalty rows come from the training $A=80$ system;
no $A=140$ path is used. At the released final MAP solution
the three rows remain active, with
$\Delta E=0.6351$, $0.4898$, and $1.6561\MeV$. Thus
$\varepsilon=2\MeV$ is a soft target margin rather than a guaranteed lower
bound. This numerical fact is also why these terms must not be described as
hard stability constraints.

With the selected hyperparameters and final active penalty rows held fixed,
let $H$ be the regularized normal matrix of the final quadratic subproblem in
the full $\bm\vartheta$ space, including the active penalty rows, and let $G$
contain the exact energy-equality rows. The inverse-Hessian Gaussian
approximation, projected onto the equality subspace, is
\begin{equation}
 \Sigma=H^{-1}-H^{-1}G^{\rm T}
 (GH^{-1}G^{\rm T})^{-1}GH^{-1},
\label{eq:gpcovariance}
\end{equation}
where the superscript $\mathrm T$ denotes transpose. This subtraction
projects $H^{-1}$ onto coefficient variations that satisfy the exact linear
equalities. It is a local Gaussian approximation to the final fixed-active-set
quadratic subproblem, not a posterior truncated or conditioned by exact
inequalities.

For an affine prediction
$\chi(r)=\chi^{(0)}(r)+\bm a(r)^{\rm T}\bm\vartheta$, the propagated variance
is
\begin{align}
 \sigma_\chi^2(r)&=\bm a(r)^{\rm T}\Sigma\bm a(r),\nonumber\\
 \sigma_{\rm NLF}(r)&\simeq
 \left|\frac{-2\chi(r)}{[1+\chi^2(r)]^2}\right|\sigma_\chi(r),
\label{eq:gppropagation}
\end{align}
where the second expression is first-order propagation through
${\rm NLF}=1/(1+\chi^2)$. The same $\Sigma$ is propagated to $\delta g(y)$
and to the fixed-density
$A=140$ response-ratio prediction shown in Fig.~\ref{fig:gp}. It does not
include hyperparameter uncertainty, uncertainty about which inequalities are
active, nonlinear
propagation through ITE, or model discrepancy. The conditional $2\sigma$
coverage is $10/80=0.125$ on the radial mask stated in
Sec.~\ref{sec:discussion}. The full finite-grid evidence selection, converged
MAP fit, conditional covariance, and coverage diagnostics are reproduced by a
single self-contained script (\texttt{gp\_fit.py}) with its staged training
inputs in the accompanying archive; it regenerates the selected hyperparameters
$(\ell_{\rm GP},\sigma_f,\kappa)=(1.5,0.3,300)$, the local-correction shift $s=-8.32492$, the
posterior mean and $2\sigma$ band, and the $A=140$ holdout coverage.

\end{document}